\documentclass[submission, Phys]{SciPost}
\usepackage{hyperref}
\def\doi{http://dx.doi.org/}
\usepackage{amsmath,braket,mathdots,mathtools,amssymb,xcolor,calligra,color}
\usepackage{amsthm} 
\usepackage{bbm}  
\usepackage{fancyhdr}
\usepackage{float}
\usepackage{authblk}
\usepackage{color}
\usepackage{authblk}
\usepackage{cite} 

\usepackage{tikz}
\usetikzlibrary{spy}
\usetikzlibrary{patterns}
\usetikzlibrary{calc,arrows,positioning}
\usetikzlibrary{arrows,shadows}
\usetikzlibrary{decorations.pathmorphing}	
\usetikzlibrary{decorations.markings}
\usepackage{pgfplots}

\usepackage{pgfplotstable}
\usepackage{filecontents}
\usepackage[utf8]{inputenc}

\newcommand{\be}{\begin{equation}}
\newcommand{\ee}{\end{equation}}
\newcommand{\bea}{\begin{eqnarray}}
\newcommand{\eea}{\end{eqnarray}}

\def\XXint#1#2#3{{\setbox0=\hbox{$#1{#2#3}{\int}$}
     \vcenter{\hbox{$#2#3$}}\kern-.5\wd0}}

\def\fr#1{(\ref{#1})}

\let\OLDthebibliography\thebibliography
\renewcommand\thebibliography[1]{
  \OLDthebibliography{#1}
  \setlength{\parskip}{0pt}
  \setlength{\itemsep}{0pt plus 0.3ex}
}

\newcommand{\sign}{\,\text{sgn}\,}
\newcommand{\Tr}{\,\text{tr}\,}

\newcommand{\D}[1]{\text{d}#1}
\newcommand{\ld}{\underset{\text{l.d.}}{\sim}}

\begin{document}
\begin{center}
{\Large\bf Low-density limit of dynamical correlations in the Lieb-Liniger model}
\end{center}
\begin{center}
Etienne Granet\textsuperscript{1$\star$},
\end{center}
\begin{center}
{\bf 1} The Rudolf Peierls Centre for Theoretical Physics, Oxford
University, Oxford OX1 3PU, UK\\
${}^\star$ {\small \sf etienne.granet@physics.ox.ac.uk}
\end{center}
\date{\today}

\section*{Abstract}
{\bf 
We derive explicit expressions for dynamical correlations of the field and density operators in the Lieb-Liniger model, within an arbitrary eigenstate with a small particle density ${\cal D}$. They are valid for all space and time and any interaction strength $c>0$, and are the leading order of an expansion in ${\cal D}$. This expansion is obtained by writing the correlation functions as sums over form factors when formally decomposed into partial fractions.

}
%

\renewcommand\Affilfont{\fontsize{9}{10.8}\itshape}

\tableofcontents
\section{Introduction}
The Lieb-Liniger model is a key paradigm of many-particle systems\cite{LiebLiniger63a,Brezin64,korepin}. In the repulsive regime, it is  considered as one of the simplest interacting quantum integrable model, for having the simplifying feature of involving only real rapidities\cite{Lieb63,YangYang69}. The main objects of interest are the correlation functions of local observables in the thermodynamic limit, that are the macroscopic output of the model resulting from the short-range interactions between the bosons. Moreover their Fourier transform is directly measurable in cold-atoms experiments \cite{naegerl15,Bouchoule16,Fabbri15}. Although exactly solvable, the computation of correlation functions in quantum integrable models is a notoriously difficult problem. In particular the computation of dynamical correlations for an arbitrary interaction strength $c$ and within arbitrary eigenstate is an open problem. This problem is relevant in a number of situations: for example both correlations at finite temperature and in the steady state reached after a quantum quench can be formulated as correlation functions within one generic averaging eigenstate\cite{YangYang69,CE13,C16,NPC15,B14}.

Special cases of this problem were studied and solved in the past. The first of these special cases is the impenetrable bosons limit $c\to\infty$, that can be reformulated in terms of free fermions \cite{girardeau}. Here, both because of the free fermionic nature of the model and the simple structure of the form factor\footnote{It is not known whether the transverse field Ising model (TFIM) could be fully treated along the same lines because of its more complex form factors, although also equivalent to free fermions \cite{GFE20}.}, the Lehmann representation of dynamical correlations can be fully resummed into a Fredholm determinant\cite{KS90,slavnov90,izergin87,kojima97} and its asymptotics extracted with differential equations \cite{IIKS90,IIKV92}. Another important special case is the ground state static correlation at finite $c$, that was treated within the Algebraic Bethe ansatz framework\cite{kitanineetalg} and led to the first ab initio calculation of the critical exponents previously predicted by Conformal Field Theory and Luttinger Liquid theory\cite{BIK86,IKR87,haldane,cazalilla}. This approach was then generalized to static correlations within arbitrary eigenstates \cite{kozlowskimailletslvanov,kozlowskimailletslvanov2}. The particular case of static correlators within thermal eigenstates was also studied with quantum transfer matrix methods\cite{suzuki85,klumper92,patuklumper}. The full asymptotics of ground state dynamical correlations were derived in \cite{kitanineetcformfactor,kozlowskimaillet,kozlowski4} from form factor expansions, confirming predictions of Non-Linear Luttinger Liquid theory\cite{IG08,PAW09,ISG12,P12,shashipanfilcaux,Price17}.

%

Progress on the general case of dynamical correlations within arbitrary states at finite $c$ has been much more limited. This calculation is very different from the special cases of zero-temperature or static cases and poses important problems. The successful methods for static correlations such as the algebraic Bethe ansatz approach or the quantum transfer matrix methods do not apply to the dynamical case, and the form factor expansion used to compute ground state dynamical correlations  relied on combinatorial identities that are usable for zero-entropy states only. This general case has however been studied through several approaches. Firstly, Generalized HydroDynamics (GHD) provide predictions for the leading asymptotics of the dynamical correlations of conserved charges such as the density within arbitrary macrostates\cite{CADY16,BCDF16,Doyon18}. However the approach cannot a priori be applied to compute the next corrections, restricting the dynamical structure factor to small frequency and momentum, and does not apply neither to semi-local operators such as the field correlations. Secondly, numerical summations of dominant form factors proved very efficient and permitted to obtain numerical estimations of the dynamical structure factor on the full plane \cite{cauxcalabreseslavnov,PC14,CC06}. On the Bethe-ansatz calculations side, one-point functions within arbitrary eigenstates were studied in \cite{leclairmussardo,saleur,kmt,pozsgay11,pozsgay12,negrosmirovn,bastianellopiroli,bastianellopirolicalabrese}. An approach based on thermodynamic form factors was started \cite{deNP15,deNP16,DNP18,panfil20,cortescuberopanfil1,cortescuberopanfil2}, but still involves non-integrable singularities and so requires a particular understanding of this feature. A regularized form factor expansion was derived for the XXZ spin chain in \cite{kozlowski1}. In \cite{granetessler20} was computed the full spectral sum at order $c^{-2}$ for all root densities and all momentum and frequency, involving one- and two-particle-hole excitations. It showed the necessity for a fine-tuned treatment of the non-integrable singularities that is crucial for e.g. detailed balance to be satisfied in thermal states.

The objective of this paper is to derive the full dynamical correlations for all space and time and all interaction strength $c$, in the limit where the particle density ${\cal D}$ of the averaging state becomes small. This low-density limit is defined in terms of a partial fraction decomposition of the form factor, initially introduced in \cite{GFE20} in the TFIM that can be reformulated in terms of free fermions. This decomposition naturally organizes the spectral sum as an expansion in the particle density ${\cal D}$ of the averaging state, and the low-density limit is defined as the leading term of this expansion. How small ${\cal D}$ has to be for this low-density limit to be accurate can only be determined by computing the next terms of the expansion.

This result provides another limiting case where the initial problem becomes solvable, namely the computation of dynamical correlations for all space and time and arbitrary $c$ within finite-entropy macrostates. Moreover the framework also enables to compute the subleading corrections in the particle density, as was explicitly shown in \cite{GFE20} for the Transverse Field Ising Model. The computation of these subleading corrections in the interacting case however comes with higher technical difficulties and should be the object of further work. Finally, this low-density limit calculation sheds light on the structure of the spectral sum and the nature of the states contributing in the thermodynamic limit.

In Section \ref{sec1} we introduce the Lieb-Liniger model and recall known results on its form factors. In Section \ref{ldd} we define what is meant by low-density limit of the dynamical correlations, taking the field correlations as an example. In Section \ref{fieldsection} we compute the low-density limit of the field two-point function \eqref{field}, and in Section \ref{densitysection} the low-density limit of the density two-point function \eqref{densityde}. In Section \ref{comments} we comment on the expressions obtained and the nature of the states that contribute to them.

\section{\texorpdfstring{Lieb-Liniger model}{Lg}}
\label{sec1}
\subsection {Definition}

The Lieb-Liniger model \cite{LiebLiniger63a} is a non-relativistic
quantum field theory model with Hamiltonian
\begin{equation}
H=\int_0^L dx\left[-\psi^\dagger(x)\frac{d^2}{dx^2}\psi(x)+c\psi^\dagger(x)\psi^\dagger(x)\psi(x)\psi(x)
\right]\,,
\label{HLL}
\end{equation}
where the canonical Bose field $\psi(x)$ satisfies equal-time
commutation relations
\begin{equation}
[\psi(x),\psi^\dagger(y)]=\delta(x-y)\,.
\end{equation}
We will impose periodic boundary
  conditions.
For later convenience we define the time-$t$ evolved version of the field $\psi(x,t)=e^{iHt}\psi(x)e^{-iHt}$. We also define the density operator at position $x$
\begin{equation}
\sigma(x)=\psi^\dagger(x)\psi(x)\,,
\end{equation}
and its time-$t$ evolved version $\sigma(x,t)=e^{iHt}\sigma(x)e^{-iHt}$.

\subsection {The Bethe ansatz solution}
\subsubsection{The spectrum}
The Lieb-Liniger model is solvable by the Bethe ansatz: an eigenstate $|\pmb{\lambda}\rangle$ with $N$ bosons can be written as
\begin{equation}\label{lambdastate}
|\pmb{\lambda}\rangle=B(\lambda_1)...B(\lambda_N)|0\rangle\,,
\end{equation}
with the $B(\lambda)$'s some creation operators, $|0\rangle$ the pseudo-vacuum and the $\lambda_i$'s some rapidities that satisfy the following set of 'Bethe equations'
\begin{equation}
e^{iL\lambda_k}=\prod_{\substack{j=1\\j\neq k}}^N\frac{\lambda_k-\lambda_j+ic}{\lambda_k-\lambda_j-ic}\,,
\quad k=1,\dots, N.
\end{equation}
The
energy $E$ and the momentum $P$ of this state read 
\begin{equation}
E(\pmb{\lambda})=\sum_{i=1}^N \lambda_i^2\,,\qquad P(\pmb{\lambda})=\sum_{i=1}^N\lambda_i\,.
\end{equation}
It is convenient to express the Bethe equations in logarithmic form
\begin{equation}
\label{belog}
\frac{\lambda_k}{2\pi}=\frac{I_k}{L}-\frac{1}{L}\sum_{j=1}^N \frac{1}{\pi}\arctan \frac{\lambda_k-\lambda_j}{c}\,,
\end{equation}
with $I_k$ an integer if $N$ is odd, a half-integer if $N$ is even. For $c>0$ all the solutions to this equation are real \cite{korepin}. We will denote
\begin{equation}
{\cal D}=\frac{N}{L}\,,
\end{equation}
the particle density of the eigenstate $|\pmb{\lambda}\rangle$.

\subsubsection{The field form factors}
Our aim is to calculate correlation functions in an eigenstate $|\pmb{\lambda}\rangle$ of the Hamiltonian \eqref{HLL}, at low particle density ${\cal D}$. We will focus on the two-point function of the field operator
\begin{equation}
  \left\langle \psi^\dagger( x,t) \psi ( 0,0) \right\rangle =\frac {\left\langle \pmb{\lambda} \left|  \psi^\dagger( x,t) \psi ( 0,0) \right| \pmb{\lambda} \right\rangle } {\left\langle \pmb{\lambda}  |\pmb{\lambda}  \right\rangle }\,,
\end{equation}
and the two-point function of the density operator
\begin{equation}
  \left\langle \sigma( x,t) \sigma ( 0,0) \right\rangle =\frac {\left\langle \pmb{\lambda} \left| \sigma( x,t) \sigma ( 0,0) \right| \pmb{\lambda} \right\rangle } {\left\langle \pmb{\lambda}  |\pmb{\lambda}  \right\rangle }\,.
\end{equation}
Our strategy is to use a Lehman representation in terms of energy eigenstates
$|\pmb{\mu}\rangle =|\mu_1,...,\mu_{N'}\rangle$, where
$\{\mu_1,\dots,\mu_{N'}\}$ are solutions to the Bethe equations \fr{belog} to rewrite the correlation functions as sums of form factors over the full spectrum. For the two-point function of a generic operator ${\cal O}$ this representation reads
\begin{equation}
\label{bigsumfield}
\begin{aligned}
  \left\langle  {\cal O}^\dagger( x,t){\cal O}( 0,0)  \right\rangle &=\sum _{ \pmb{\mu}}\frac {\left| \left\langle \pmb{\mu} |{\cal O}( 0) |\pmb{\lambda}\right\rangle \right| ^{2}} {\left\langle \pmb{\lambda} \left| \pmb{\lambda} \right\rangle \left\langle \pmb{\mu}\right| \pmb{\mu}\right\rangle }e^{it\left( E\left( \pmb{\lambda}\right) -E\left( \pmb{\mu} \right) \right) +ix\left( P\left( \pmb{\mu}\right) -P\left( \pmb{\lambda}\right) \right) }\,.
\end{aligned}
\end{equation}

The (normalized) form factors of the field and density operators between two Bethe
states $|\pmb{\lambda}\rangle,|\pmb{\mu}\rangle$ with respective numbers of Bethe roots $N,N'$ have been computed previously \cite{Korepin82,Slavnov89,Slavnov90,KorepinSlavnov99,Oota04,KozlowskiForm11}. 

In the case of the field operator, it reads
\begin{equation}\label{FF}
\begin{aligned}
&\frac { \left\langle \pmb{\mu} |\psi ( 0) |\pmb{\lambda}\right\rangle  } {\sqrt{\left\langle \pmb{\lambda} \left| \pmb{\lambda} \right\rangle \left\langle \pmb{\mu}\right| \pmb{\mu}\right\rangle }}=\delta_{N,N'+1}\frac{i^{N+1}(-1)^{N(N-1)/2}}{L^{N-1/2}\sqrt{\mathcal{N}_{\pmb{\lambda}}\mathcal{N}_{\pmb{\mu}}}}\frac{\prod_{i< j}|\lambda_i-\lambda_j|\prod_{i< j}|\mu_i-\mu_j|}{\prod_{i, j}(\mu_j-\lambda_i)}\\
&\qquad\times\sqrt{\frac{\prod_{i\neq j}(\lambda_i-\lambda_j+ic)}{\prod_{i\neq j}(\mu_i-\mu_j+ic)}}\prod_{\substack{j=1\\j\neq p,s}}^N(V_j^+-V_j^-) \,\,\underset{i,j=1,...,N}{\det}\Bigg[\delta_{ij}+U_{ij}\Bigg]\,,
\end{aligned}
\end{equation}
for any $p,s=1,...,N$. The various terms entering this expression are
\begin{equation}
V_i^\pm=\frac{\prod_{k=1}^{N-1}(\mu_k-\lambda_i\pm ic)}{\prod_{k=1}^{N}(\lambda_k-\lambda_i\pm ic)}\,,
\end{equation}
and the $N\times N$ matrix
\begin{equation}
U_{jk}=\frac{i}{V_j^+-V_j^-}\left[\frac{2c}{c^2+(\lambda_j-\lambda_k)^2}-\frac{4c^2}{(c^2+(\lambda_p-\lambda_k)^2)(c^2+(\lambda_s-\lambda_j)^2)}\right]\frac{\prod_{m=1}^{N-1}(\mu_m-\lambda_j)}{\prod_{m\neq j}(\lambda_m-\lambda_j)}\,,
\end{equation}
and finally
\begin{equation}
\label{norm}
\mathcal{N}_{\pmb{\lambda}}=\det G(\pmb{\lambda})\,,
\end{equation}
with the Gaudin matrix \cite{Gaudin71}
\begin{equation}\label{gaudin}
G_{ij}(\pmb{\lambda})= \delta_{ij} \left(1+\frac{1}{L}\sum_{k=1}^N \frac{2c}{c^2+(\lambda_i-\lambda_k)^2}\right)-\frac{1}{L}\frac{2c}{c^2+(\lambda_i-\lambda_j)^2}\,.
\end{equation}
The form factor of the density operator reads
\begin{equation}\label{desnityff}
\begin{aligned}
&\frac { \left\langle \pmb{\mu} |\sigma ( 0) |\pmb{\lambda}\right\rangle  } {\sqrt{\left\langle \pmb{\lambda} \left| \pmb{\lambda} \right\rangle \left\langle \pmb{\mu}\right| \pmb{\mu}\right\rangle }}=\delta_{N,N'}\frac{i^{N+1}(-1)^{N(N-1)/2}(\sum_{j=1}^N \lambda_j-\mu_j)}{L^{N}\sqrt{\mathcal{N}_{\pmb{\lambda}}\mathcal{N}_{\pmb{\mu}}}}\frac{\prod_{i< j}|\lambda_i-\lambda_j|\prod_{i< j}|\mu_i-\mu_j|}{\prod_{i, j}(\mu_j-\lambda_i)}\\
&\qquad\times\sqrt{\prod_{i, j}\frac{\lambda_i-\lambda_j+ic}{\mu_i-\mu_j+ic}}\prod_{j\neq p}(V_j^+-V_j^-) \,\,\underset{i,j=1,...,N}{\det}\Bigg[\delta_{ij}+U'_{ij}\Bigg]\,,
\end{aligned}
\end{equation}
for any $p=1,...,N$, with
\begin{equation}
U'_{jk}=i\frac{\mu_j-\lambda_j}{V_j^+-V_j^-}\left[\frac{2c}{(\lambda_j-\lambda_k)^2+c^2}-\frac{2c}{(\lambda_p-\lambda_k)^2+c^2}\right]\prod_{m\neq j}\frac{\mu_m-\lambda_j}{\lambda_m-\lambda_j}\,,
\end{equation}
and with now
\begin{equation}
V_j^\pm=\frac{\prod_{k=1}^{N}(\mu_k-\lambda_j\pm ic)}{\prod_{k=1}^{N}(\lambda_k-\lambda_j\pm ic)}\,.
\end{equation}
\subsubsection{Root densities}
In the thermodynamic limit, any sum of a non-singular function over the Bethe roots can be expressed in terms of a \textit{root density} that characterizes a macrostate as far as such quantities are concerned
\begin{equation}
\underset{L\to\infty}{\lim}\, \frac{1}{L^n}\sum_{i_1,...,i_n}f(\lambda_{i_1},...,\lambda_{i_n})=\int_{-\infty}^\infty \dots\int_{-\infty}^\infty f(\lambda_1,...,\lambda_n)\rho(\lambda_1)\dots\rho(\lambda_n)\D{\lambda_1}\dots\D{\lambda_n}\,.
\end{equation}
However if the function is singular the result will in general depend on the representative state of the macrostate, see \cite{granetessler20}.
It is customary to introduce the hole density $\rho_h(\lambda)$ defined by
\begin{equation}
\label{vartheta}
\rho(\lambda)+\rho_h(\lambda)=\frac{1}{2\pi}+\frac{1}{2\pi}\int_{-\infty}^\infty \frac{2c}{c^2+(\lambda-\mu)^2}\rho(\mu)\D{\mu}\,.
\end{equation}
In the rest of the paper, $\rho(\lambda)$ denotes the root density of the averaging eigenstate \eqref{lambdastate} in which we compute the two-point functions of interest. We emphasize that this root density is considered as a \textit{given data} of the problem. It means we do not consider the low-density expansion of the solutions to the Thermodynamic Bethe Ansatz equations describing e.g. finite-temperature states.

\section {Definition of the low-density limit \label{ldd}}
The purpose of this section is to define what is meant by the \textit{low density limit} of correlation functions. It is defined as the \textit{leading order of an expansion} in ${\cal D}$, obtained by decomposing the form factor in partial fractions. As a consequence it is an expression valid for all $x,t$ and $c$, that becomes closer to the dynamical correlations as the particle density ${\cal D}$ of the averaging state becomes smaller.\\

This definition requires some technicalities, but is rigorous and allows for a computation of the next orders, as shown in \cite{GFE20} for the TFIM that can be reformulated in terms of free fermions. However, it a priori lacks some intuitive picture. For that reason we provide an interpretation of this low-density limit so defined as a Lehmann representation in terms of the \textit{low density limit of the form factor}. The reasoning is here rather different, and consists in first approximating the form factor by the thermodynamic limit value they take when one of the two states is a dilute state, i.e. such that for any pair $i,j$ we have  $L(\lambda_i-\lambda_j)\to\infty$. In this limit, the spectral sum of the dynamical correlations indeed matches the leading order of the expansion in ${\cal D}$, providing an interesting and intuitive consistency check. 

%

\subsection {Partial fraction decomposition}
\subsubsection{Recall \label{pfddefsec}}
We recall that the partial fraction decomposition (PFD) of a ratio of two polynomials $\frac{P(X)}{\prod_{i=1}^n(X-x_i)^{a_i}}$ with distinct $x_i$'s is the writing
\begin{equation}\label{pfddef}
\frac{P(X)}{\prod_{i=1}^n(X-x_i)^{a_i}}=P_0(X)+\sum_{i=1}^n \sum_{\nu=1}^{a_i}\frac{B_{i,\nu}}{(X-x_i)^\nu}\,,
\end{equation}
with $P_0(X)$ a polynomial, and $B_{i,\nu}$ coefficients given by
\begin{equation}
B_{i,\nu}=\frac{1}{(a_i-\nu)!}(\tfrac{d}{dX})^{a_i-\nu}[(X-x_i)^{a_i}P(X)]|_{X=x_i}\,.
\end{equation}
The polynomial $P_0(X)$ can be determined by studying e.g. the large $X$ behaviour of the ratio of the two polynomials on the left-hand side of \eqref{pfddef}.
\subsubsection{The poles of the normalized form factor}
We consider $ \pmb{\lambda}  $ and $ \pmb{\mu}  $ two sets of respectively $N$ and $N-1$ rapidities, that do not necessarily satisfy the Bethe equations. We  define the reduced form factor $F_\psi(\pmb{\lambda},\pmb{\mu})$ by 
\begin{equation}\label{pfd}
\begin{aligned}
\frac {\left| \left\langle \pmb{\mu} |\psi \left( 0\right) |\pmb{\lambda}\right\rangle \right| ^{2}} {\left\langle \pmb{\lambda} \left| \pmb{\lambda} \right\rangle \left\langle \pmb{\mu}\right| \pmb{\mu}\right\rangle }= \frac{F_\psi(\pmb{\lambda},\pmb{\mu})}{\mathcal{N}_{\pmb{\lambda}}\mathcal{N}_{\pmb{\mu}}L^{2N-1}} \,,
\end{aligned}
\end{equation}
and would like to apply a PFD to $F_\psi(\pmb{\lambda},\pmb{\mu})$ seen as a function of each of the $\mu_i$'s successively. From \eqref{FF}, we find that in $F_\psi(\pmb{\lambda},\pmb{\mu})$ there are two types of poles for a $\mu_i$ at fixed other $\mu_j$'s
\begin{enumerate}
\item Double poles in $\lambda_j$ for all $j$
\item Simple poles in $\mu_j \pm ic$ for all $j\neq i$
\end{enumerate}

We note that the second type of pole is never attained when the $\mu_i$'s satisfy the Bethe equations, since the rapidities are real in this case. However, to perform the PFD of the form factor one has to consider the functional dependence of $F_\psi(\pmb{\lambda},\pmb{\mu})$ on the parameters $\mu_i$'s, irrespectively of the fact that the PFD will be evaluated for real $\mu_i$'s in the end, so these poles have to be taken into account in the PFD indeed.

\subsubsection{The shape of the PFD of the reduced form factor}
The reduced form factor $F_\psi(\pmb{\lambda},\pmb{\mu})$, seen as a function of $\mu_1$, is a ratio of two polynomials with double poles in each of the $\lambda_i$'s and simple poles in $\mu_j\pm ic$, so one can apply the decomposition written in Section \ref{pfddefsec}. Since the reduced form factor goes to zero when $\mu_1\to\infty$, we have $P_0(X)=0$ and so one can write
\begin{equation}\label{154}
F_\psi(\pmb{\lambda},\pmb{\mu})=h_{\mu_1}(\mu_2,...,\mu_{N-1})+\sum_{i=1}^N \sum_{\nu=1}^{2}\frac{B_{i,\nu}(\mu_2,...,\mu_{N-1})}{(\mu_1-\lambda_i)^\nu}\,,
\end{equation}
with
\begin{equation}
h_{\mu_1}(\mu_2,...,\mu_{N-1})=\sum_{i=2}^{N-1} \frac{C^+_{i}(\mu_2,...,\mu_{N-1})}{\mu_1-\mu_i+ic}+\sum_{i=2}^{N-1} \frac{C^-_{i}(\mu_2,...,\mu_{N-1})}{\mu_1-\mu_i-ic}\,,
\end{equation}
where $B_{i,\nu}(\mu_2,...,\mu_{N-1})$, $C^\pm_{i}(\mu_2,...,\mu_{N-1})$ are 'coefficients' independent of $\mu_1$, but that still possess a dependence in the remaining $\mu_k$'s. This writing is nothing but the PFD of $F_\psi(\pmb{\lambda},\pmb{\mu})$ seen as a function of $\mu_1$. The coefficients $B_{i,\nu}$ depend both on $i$ the index of the rapidity $\lambda_i$ at which $\mu_1$ has a pole, and on $\nu$ the order of the pole. The coefficients $C_{i}^\pm$ depend as well on $i$ the index of the rapidity such that $\mu_1$ has a pole at $\mu_i\pm ic$, but since these poles are always simple there is no dependence on $\nu$. There is however an additional upper index $\pm$ to distinguish the cases $\pm ic$. These coefficients have the same pole structure as $F_\psi(\pmb{\lambda},\pmb{\mu})$, except that each $\mu_i$ for $i\neq 1$ does not anymore have a pole in $\mu_1\pm ic$. We note that by definition, the function $h_{\mu_1}(\mu_2,...,\mu_{N-1})$ is a function of $\mu_1$ that has no poles in $\mu_1$ when all the $\mu_j$'s are real. 

We now apply the same procedure to $B_{i,\nu}(\mu_2,...,\mu_{N-1})$ and $h_{\mu_1}(\mu_2,...,\mu_{N-1})$ with respect to $\mu_2$. It yields
\begin{equation}
\begin{aligned}
h_{\mu_1}(\mu_2,...,\mu_{N-1})&=h_{\mu_1,\mu_2}(\mu_3,...,\mu_{N-1})+\sum_{i=1}^N \sum_{\nu=1}^{2}\frac{B_{i,\nu| \mu_1}(\mu_3,...,\mu_{N-1})}{(\mu_2-\lambda_i)^\nu}\\
B_{i,\nu}(\mu_2,...,\mu_{N-1})&=k_{\mu_2}(\mu_3,...,\mu_{N-1})+\sum_{i=1}^N \sum_{\nu'=1}^{2}\frac{B_{i,\nu,j,\nu'}(\mu_3,...,\mu_{N-1})}{(\mu_2-\lambda_i)^{\nu'}}\,,
\end{aligned}
\end{equation}
where $B_{i,\nu,j,\nu'}(\mu_3,...,\mu_{N-1})$ is a 'coefficient' independent of $\mu_1$ and $\mu_2$, and $B_{i,\nu| \mu_1}(\mu_3,...,\mu_{N-1})$ a  function of $\mu_1$ without singularities in real $\mu_1$, and independent of $\mu_2$. The function $h_{\mu_1,\mu_2}(\mu_3,...,\mu_{N-1})$ is a  function of $\mu_1,\mu_2$ without singularities in real $\mu_1,\mu_2$, and $k_{\mu_2}(\mu_3,...,\mu_{N-1})$ a function of $\mu_2$ without singularities in real $\mu_2$, and independent of $\mu_1$. Plugged into \eqref{154}, it follows that $F_{\psi}(\pmb{\lambda},\pmb{\mu})$ can be written
\begin{equation}
F_{\psi}(\pmb{\lambda},\pmb{\mu})=\sum_{\nu_1=0}^2\sum_{\nu_2=0}^2 \sum_{\substack{f_1=1\\ \text{if }\nu_1>0}}^N \sum_{\substack{f_2=1\\ \text{if }\nu_2>0}}^N\frac{a(\mu_1,\mu_2|\mu_3,...,\mu_{N-1})}{(\mu_1-\lambda_{f_1})^{\nu_1}(\mu_2-\lambda_{f_2})^{\nu_2}}\,,
\end{equation}
where $a$ is a coefficient that depends respectively on $\mu_1,\mu_2$  \textit{only if} $\nu_1=0,\nu_2=0$. The sums over $f_1,f_2$ are present only if $\nu_1>0,\nu_2>0$ (if $\nu_i=0$ then $f_i$ does not appear in the summand and by convention the sum symbol is removed).


Proceeding recursively by applying a PFD to $a(\mu_1,\mu_2|\mu_3,...,\mu_{N-1})$ with respect to $\mu_3$, and then to $\mu_4$ etc, one obtains the writing
\begin{equation}\label{pfd}
F_\psi(\pmb{\lambda},\pmb{\mu})=\sum_{\substack{\nu_1,...,\nu_{N-1}\\\in\{0,1,2\}}}\sum_{f}\frac{A(\pmb{\lambda},\pmb{\mu},\{\nu\},f)}{\prod_{\substack{i=1\\ \nu_i>0}}^{N-1}(\mu_i-\lambda_{f(i)})^{\nu_i}}\,.
\end{equation}
The sum over $f$ denotes a sum over all the functions defined on the points $i\in\{1,...,N-1\}$ where $\nu_i>0$ and that take values in $\{1,...,N\}$, namely over all the functions $f$ such that
\begin{equation}
f:\{i\in\{1,...,N-1\} | \nu_i>0 \}\to \{1,...,N\}\,.
\end{equation}
The coefficients $A(\pmb{\lambda},\pmb{\mu},\{\nu\},f)$ crucially do not depend on any $\mu_i$ whenever $\nu_i>0$, and are bounded regular functions of real $\mu_i$ when $\nu_i=0$. 

\subsubsection{Computing the coefficients}
In the special case where all $\nu_i>0$, the coefficients $A(\pmb{\lambda},\pmb{\mu},\{\nu\},f)=A(\pmb{\lambda},\{\nu\},f)$ do not depend on any $\mu_i$'s and can be computed according to the formula
\begin{equation}
\label{formulA}
A(\pmb{\lambda},\{\nu\},f)=\prod_{i=1}^{N-1}\left[\left(\frac{d}{d\mu_i}\right)^{2-\nu_i} (\mu_i-\lambda_{f(i)})^2 \right]F_\psi(\pmb{\lambda},\pmb{\mu})|_{\mu_i=\lambda_{f(i)}}\,.
\end{equation}
An example of use of this formula is provided below in Section \ref{partyop1}.

If now there is a subset $K\subset\{1,...,N-1\}$ such that $\nu_i=0$ for $i\in K$, one first defines the following function of the $\mu_i$'s for $i\in K$
\begin{equation}
\label{formulA2}
\bar{A}(\{\mu_i\}_{i\in K}|\pmb{\lambda},\{\nu\},f)=\prod_{\substack{i=1\\ i\notin K}}^{N-1}\left[\left(\frac{d}{d\mu_i}\right)^{2-\nu_i} (\mu_i-\lambda_{f(i)})^2 \right]F_\psi(\pmb{\lambda},\pmb{\mu})|_{\mu_i=\lambda_{f(i)},\, i\notin K}\,.
\end{equation}
The function $\bar{A}(\{\mu_i\}_{i\in K}|\pmb{\lambda},\{\nu\},f)$ still has poles in $\mu_i$ for $i\in K$ since it also contains all the cases when $\nu_i>0$. To compute the coefficient $A(\pmb{\lambda},\pmb{\mu},\{\nu\},f)$ one has to remove from this function any pole in real $\mu_i$ for $i\in K$. An example of use of this formula is provided below in Section \ref{partyop12}.

Although these formulas may not look very explicit, they are still of practical use to compute the simplest terms in the PFD, as we will see below.\\

The functions $f$ over which we sum in \eqref{pfd} are actually rather constrained.

Let us first consider the case where $\nu_i=\nu_j=2$ and $f(i)=f(j)$. Then from formula \eqref{formulA} (if all $\nu_k>0$ are non-zero) or formula \eqref{formulA2} (if there is a $\nu_k=0$), since there is a zero $(\mu_i-\mu_j)^2$ in the form factor without derivatives with respect to $\mu_i$ or $\mu_j$, the coefficient $A$ vanishes. Hence one can constrain $f$ to be such that $f(i)\neq f(j)$ if $\nu_i=\nu_j=2$.

Secondly, let us consider the case where we have $\nu_i=2$ and $\nu_j=1$, and $f(i)=f(j)$. There is still a double zero $(\mu_i-\mu_j)^2$ in the form factor, without derivative with respect to $\mu_i$ and with only one derivative with respect to $\mu_j$. Hence in this case the coefficient $A$ also vanishes. It follows that we can impose that whenever $\nu_i=2$ at a point $i$, the function $f$ takes only once the value $f(i)$.

Thirdly, let us now assume that $k$ indices have $\nu_i=1$ and all take the same value through $f$. Then in the form factor there is a zero of order $k(k-1)$, but with only $k$ derivatives applying on this zero in formulas \eqref{formulA} and \eqref{formulA2}. For the coefficient $A$ not to vanish, the only possibility is $k=2$. Hence one can impose that in any case $f$ can take at most twice the same value.

Thus we conclude that one can impose in \eqref{pfd} the two following constrains: (i) that $f(i)\neq f(j)$ whenever $\nu_i=2$ or $\nu_j=2$, and (ii) that $f$ can take at most twice the same value.\\

\subsection {A density expansion}

Let us now rewrite the Lehmann representation \eqref{bigsumfield} in the following way. Instead of summing over the Bethe roots $\mu_i$, we sum over their Bethe numbers $J_i$, and trade the ordering of the Bethe numbers for a non-ordered sum with a $\frac{1}{(N-1)!}$ factor. Whenever two Bethe numbers coincide, the form factor is zero so that the two representations are indeed equivalent. Using \eqref{pfd}, we obtain
\begin{equation}\label{pfdbeg}
\begin{aligned}
 \left\langle  \psi^\dagger\left( x,t\right) \psi \left( 0,0\right)  \right\rangle =&\frac{1}{L^{2N-1}(N-1)!}\\
 &\sum_{\{\nu\},f}\sum_{J_1,...,J_{N-1}}\frac{A(\pmb{\lambda},\pmb{\mu},\{\nu\},f)}{\mathcal{N}_{\pmb{\lambda}}\mathcal{N}_{\pmb{\mu}}}\frac{e^{it\left( E( \pmb{\lambda}) -E( \pmb{\mu}) \right) +ix\left( P( \pmb{\mu}) -P( \pmb{\lambda}) \right) }}{\prod_{i=1}^{N-1}(\mu_i-\lambda_{f(i)})^{\nu_i}}\,.
\end{aligned}
\end{equation}
The sum over $J_1,...,J_{N-1}$ is invariant under any change $\tilde{f}=f \circ (i \,j)$ with $(i \,j)$ the permutation of indices $i,j$, whenever $\nu_i=\nu_j>0$ and $f(i)$ and $f(j)$ are attained the same number of times by $f$. Hence this sum only depends on the \textit{set of points attained a given number of times} by $f$, not on the particular realization of the function $f$.

In the following we will denote the number of elements of a set $E$ by
\begin{equation}
|E|\qquad \text{or}\qquad \# E\,,
\end{equation}
according to the most readable choice in the context.

To rewrite the sum without these functions $f$, let us define $I_k$ for $k=0,1,2$ the set of points $i$ in $\{1,...,N\}$ that are attained $k$ times by $f$ from points where $\nu_j=1$, namely
\begin{equation}
\begin{aligned}
I_k&=\Bigg\{i\in\{1,...,N\} \left| \# \{j\in \{1,...,N-1\}| \nu_j=1\text{ and }f(j)=i\}=k\Bigg\}\right.\,.
\end{aligned}
\end{equation}
As a consequence the points in $\{1,...,N\}$ attained by $f$ from points where $\nu_j=2$ are $\{1,...,N\}-(I_0\cup I_1\cup I_2)$. These subsets $I_0,I_1,I_2\subset \{1,...,N\}$ have to be disjoint and to satisfy $|I_0|=|I_2|+1+p$ with $p=|\{i|\nu_i=0\}|$ the number of points with $\nu_i=0$, because  $f$ can take at most twice the same value. 

We will denote
\begin{equation}
n=|I_2|\,,\qquad m=|I_1|\,,
\end{equation}
and parametrize
\begin{equation}
\begin{aligned}
I_2&=\{j_{1},...,j_{n}\}\\
I_0&=\{j_{n+1},...,j_{2n+p+1}\}\\
I_1&=\{j_{2n+p+2},...,j_{2n+p+m+1}\}\\
\{1,...,N\}-(I_0\cup I_1 \cup I_2)&=\{j_{2n+p+m+2},...,j_{N}\}\,.
\end{aligned}
\end{equation}
When rewriting \eqref{pfdbeg} in terms of these subsets, one picks  a combinatorial factor corresponding to the number of functions $f$ with such an output. Let us determine now this combinatorial factor. Choosing the set of points where $\nu_i=0$ yields a factor ${N-1\choose p}$, those where $\nu_i=2$ a factor $(N-2n-p-m-1)!{N-1-p\choose N-2n-p-m-1}$, those attained only once by $f$ and where $\nu_i=1$ a factor $m!{2n+m\choose m}$. Finally those attained twice by $f$ yield a factor $(2n)!!n!$. Writing $(2n)!!=\tfrac{(2n)!}{n!2^n}$ yields a total combinatorial factor
\begin{equation}
\frac{(N-1)!}{2^{n}p!}\,.
\end{equation}
We conclude that we can write
\begin{equation}\label{expdensity}
\left\langle  \psi^\dagger\left( x,t\right) \psi \left( 0,0\right)  \right\rangle =\sum_{n,m,p\geq 0}S_{n,m,p}\,,
\end{equation}
with
\begin{equation}
\begin{aligned}
&S_{n,m,p}=\frac{1}{2^np!L^{2N-1}}\sum_{\substack{I_{0,1,2}\subset \{1,...,N\}\\|I_0|=n+p+1\\|I_1|=m\\|I_2|=n \\\text{all disjoint}}}\sum_{J_1,...,J_{N-1}}\frac{{\cal A}(I_0,I_1,I_2|\{\mu_i\}_{i=2n+1}^{2n+p})}{\mathcal{N}_{\pmb{\lambda}}\mathcal{N}_{\pmb{\mu}}}\\
  &\qquad\qquad\times\frac{e^{it\left( E\left( \pmb{\lambda}\right) -E\left( \pmb{\mu} \right) \right) +ix\left( P\left( \pmb{\mu}\right) -P\left( \pmb{\lambda}\right) \right) }}{\prod_{i=1}^{n}(\mu_{2i-1}-\lambda_{j_i})(\mu_{2i}-\lambda_{j_i})\prod_{i=2n+1+p}^{2n+m+p}(\mu_i-\lambda_{j_{i+1}})\prod_{i=2n+m+p+1}^{N-1}(\mu_i-\lambda_{j_{i+1}})^{2}}\,.
\end{aligned}
\end{equation}
The specific ordering of the $\mu_i$'s in this expression is irrelevant. 

Here we have ${\cal A}(I_0,I_1,I_2|\{\mu_i\}_{i=2n+1}^{2n+p})=A(\pmb{\lambda},\pmb{\mu},\{\nu\},f)$ with $\nu_i=1$ for $i=1,...,m+2n$, $\nu_i=0$ for $i=m+2n+1,...,m+2n+p$ and $\nu_i=2$ for $i=m+2n+p+1,...,N-1$, and with the function $f$ taking the values in $I_1$ over $1,...,m$, twice the values in $I_2$ over $m+1,...,m+2n$, and the values in $\{1,...,N\}-(I_0\cup I_1\cup I_2)$ over $m+2n+p+1,...,N-1$.\\

Let us now explain why the expansion \eqref{expdensity} is an expansion in the density of particles $\mathcal{D}$ of the averaging state. The term $S_{n,m,p}$ is a sum over three subsets of $\{\lambda_1,...,\lambda_N\}$ of sizes $n+p+1$, $m$ and $n$. Hence the sum of all the $S_{n,m,p}$ at fixed $2n+p+m+1\equiv \Delta$ can be written as a $\Delta$-fold sum over the Bethe roots. In the thermodynamic limit it can thus be expressed as a $\Delta$-fold integral over the root density $\rho(\lambda)$, which is thus of order ${\cal O}({\cal D}^{\Delta})$. Hence once the terms are regrouped at fixed $2n+p+m+1$, the expression \eqref{expdensity} is \textit{an expansion in the particle density} ${\cal D}$ of the averaging state. 

We note that it may be necessary indeed to sum over all the $S_{n,m,p}$'s at fixed $2n+p+m+1$ to have a finite quantity in the thermodynamic limit. This fact was  observed previously in the TFIM \cite{GFE20}.

\subsection {Definition of the low-density limit of the correlation function\label{partyop1}}
The low density limit of the dynamical correlations is defined as retaining only the leading term $S_{0,0,0}$ in \eqref{expdensity}. It is obtained with $p=0$ and $I_1=I_2=\varnothing$, and so $I_0=\{a\}$ for $a=1,...,N$. Namely, reparametrising $\pmb{\mu}=\{\mu_1,...,\mu_{a-1},\mu_{a+1},...,\mu_N\}$ for convenience
\begin{equation}\label{S00}
\begin{aligned}
  &S_{0,0,0}=\frac{1}{L}\sum_{a=1}^N\sum_{J_i,\, i\neq a}\frac{{\cal A}(\{a\},\varnothing,\varnothing|\varnothing)}{\mathcal{N}_{\pmb{\lambda}}\mathcal{N}_{\pmb{\mu}}}\frac{e^{it\left( E\left( \pmb{\lambda}\right) -E\left( \pmb{\mu} \right) \right) +ix\left( P\left( \pmb{\mu}\right) -P\left( \pmb{\lambda}\right) \right) }}{\prod_{i\neq a}L^2(\mu_i-\lambda_{i})^{2}}\,.
\end{aligned}
\end{equation}
Since all the other terms in \eqref{expdensity} have at least a multiplying factor ${\cal D}$, we have
\begin{equation}\label{psipsild2}
\begin{aligned}
  &\left\langle  \psi^\dagger\left( x,t\right) \psi \left( 0,0\right)  \right\rangle =S_{0,0,0}(1+{\cal O}({\cal D}))\,.
\end{aligned}
\end{equation}
In the rest of paper, we will use the sign $\ld$ to indicate a low-density limit. Namely
\begin{equation}
X\ld Y\qquad \text{means}\qquad X=Y(1+{\cal O}({\cal D}))\,.
\end{equation}
Using formula \eqref{formulA}, one finds
\begin{equation}
{\cal A}(\{a\},\varnothing,\varnothing|\varnothing)=\prod_{i\neq a}\frac{4c^2}{(\lambda_i-\lambda_a)^2+c^2}\,.
\end{equation}
Hence the low-density limit
\begin{equation}\label{psipsild2}
\begin{aligned}
  &\left\langle  \psi^\dagger\left( x,t\right) \psi \left( 0,0\right)  \right\rangle \ld\frac{1}{L}\sum_{a=1}^Ne^{it\lambda_a^2-ix\lambda_a}\sum_{J_i,\, i\neq a}\frac{1}{\mathcal{N}_{\pmb{\lambda}}\mathcal{N}_{\pmb{\mu}}}\prod_{i\neq a}\left(\frac{4c^2}{(\lambda_i-\lambda_a)^2+c^2}\frac{e^{it\left( \lambda_i-\mu_i \right) +ix\left( \mu_i-\lambda_i \right) }}{L^2(\mu_i-\lambda_{i})^{2}}\right)\,.
\end{aligned}
\end{equation}
A few comments are in order. Up to now, nothing has been said of the Bethe equations, and in principle the $\mu_i$'s in this expression should satisfy exactly the Bethe equations \eqref{belog}. If one wishes to determine the dynamical correlations at leading order in ${\cal D}$, there remains only the term \eqref{psipsild2} in the full expansion \eqref{expdensity}; but one can also satisfy the Bethe equations  \eqref{belog} only at leading order in ${\cal D}$, since their exact solution will involve higher orders in ${\cal D}$ that are of the same order as the terms discarded in \eqref{expdensity}. Stated differently, the leading order in ${\cal D}$ of the dynamical correlations is obtained by both retaining only \eqref{psipsild2} \textit{and} satisfying \eqref{belog} at leading order in ${\cal D}$, while the higher orders in ${\cal D}$ will require both taking into account higher terms in \eqref{expdensity}  \textit{and} satisfying \eqref{belog} at higher orders in ${\cal D}$ in \eqref{psipsild2}.

\subsection {Interpretation: low-density limit of the field form factor\label{intuitive}}
The low-density limit is defined above as the leading term in an expansion of the correlation functions obtained by decomposing the form factors in partial fractions, that turns out to be an expansion in the density of the averaging state ${\cal D}$. 
This definition allows for a systematic calculation of the next corrections in the density by taking into account more terms in \eqref{expdensity}.\\

The low-density limit of the correlation functions can however be recovered more intuitively but less rigorously with the following reasoning. If the root density $\rho(\lambda)$ of the averaging state is small, then the distance between two consecutive roots $L(\lambda_i-\lambda_j)$ is 'typically'\footnote{This cannot be true for any representative state of the density, but is true for a 'typical state' whose roots are regularly spaced according to the value of the density, see \cite{granetessler20}.} large in front of $1$, and in the limit of vanishingly low density becomes infinite. We will say that a sequence of states satisfying this property is \textit{dilute}, namely
\begin{equation}
(\pmb{\lambda}^{(L)})_{L>0}\text{ dilute}\iff \forall i\neq j,\, \underset{L\to\infty}{\lim}\, L|\lambda^{(L)}_i-\lambda^{(L)}_j|=\infty\,.
\end{equation}
For notational convenience we will drop the $L$ dependence of the sequence of states. Let us consider a dilute state $\pmb{\lambda}$ and investigate the consequences it has on the form factors between $\pmb{\lambda}$ and another state $\pmb{\mu}$.

Let us first note that because of the assumption of diluteness, a Bethe number $J_i$ of $\pmb{\mu}$ can be at a distance ${\cal O}(1)$ of at most one Bethe number $I_j$ of $\pmb{\lambda}$. We can translate this affirmation in terms of distance between rapidities with the following reasoning. Defining the counting function
\begin{equation}
\xi(\lambda)=\frac{\lambda}{2\pi}+\frac{1}{\pi}\int_{-\infty}^\infty \arctan \left(\frac{\lambda-\mu}{c}\right)\rho(\mu)\D{\mu}\,,
\end{equation}
that is such that $\xi(\lambda_i)=\frac{I_i}{L}+o(L^0)$, we have $\xi'(\lambda)=\rho(\lambda)+\rho_h(\lambda)$ that from \eqref{vartheta} satisfies
\begin{equation}
\frac{1}{2\pi}\leq \rho(\lambda)+\rho_h(\lambda) \leq \frac{1}{2\pi}+\frac{\mathcal{D}}{\pi c}\,.
\end{equation}
Hence
\begin{equation}
\frac{1}{\max (\rho+\rho_h)}|\tfrac{I_i}{L}-\tfrac{I_j}{L}|\leq |\lambda_i-\lambda_j|\leq \max \left(\frac{1}{\rho+\rho_h}\right) |\tfrac{I_i}{L}-\tfrac{I_j}{L}|\,,
\end{equation}
which shows that the distance between two rapidities is of the same order in $L$ as the distance between their Bethe numbers divided by $L$. Hence given a root $\mu_i$, the quantity $L(\mu_i-\lambda_j)$ can be of order $1$ for at most one $\lambda_j$.

Let us now denote $n_i$ the number of roots $\mu_j$'s at a distance $\mathcal{O}(L^{-1})$ from $\lambda_i$. We have
\begin{equation}
\prod_{j<k}|\mu_j-\mu_k|=\mathcal{O}(L^{-\sum_{i=1}^{N}\tfrac{n_i(n_i-1)}{2}})\,,\qquad \prod_{i,j}|\lambda_i-\mu_j|=\mathcal{O}(L^{-\sum_{i=1}^N n_i})\,.
\end{equation}
Consequently, the form factor \eqref{FF} in the low density limit is non-zero only if $n_i=1$ for all $i=1,...,N, \neq a$  for a certain $a\in\{1,...,N\}$. It means that all the roots $\lambda_i$ except one are surrounded by exactly one $\mu_j$ at a distance $\mathcal{O}(L^{-1})$. We can then re-label the roots $\pmb{\mu}=\{\mu_1,...,\mu_{a-1},\mu_{a+1},...,\mu_N\}$ so that $L(\mu_i-\lambda_i)$ is of order $1$ for all $i\neq a$. \\

Let us then investigate the value taken by the normalized form factor \eqref{FF} in this regime.

We start by evaluating the determinant $\mathcal{N}_{\pmb{\lambda}}$ in the low-density limit. We have
\begin{equation}\label{gaudin2}
G_{ij}(\pmb{\lambda})= \delta_{ij}-\frac{1}{L}\frac{2c}{c^2+(\lambda_i-\lambda_j)^2}+{\cal O}({\cal D})\,,
\end{equation}
so that
\begin{equation}\label{det1det1}
\begin{aligned}
\mathcal{N}_{\pmb{\lambda}}&=\exp \Tr \log G(\pmb{\lambda})\\
&=\exp\left( -\sum_{n=1}^\infty \frac{1}{n}\Tr g^n+{\cal O}({\cal D})\right)\\
&=1+{\cal O}({\cal D})\,,
\end{aligned}
\end{equation}
with $g_{ij}=\frac{1}{L}\frac{2c}{c^2+(\lambda_i-\lambda_j)^2}$, since $\Tr g^n={\cal O}({\cal D}^n)$. Here, we did not use the diluteness of $\pmb{\lambda}$, but only evaluated the leading order in ${\cal D}$ of the determinant.

Secondly, again because $\pmb{\lambda}$ is dilute, all the $L(\mu_i-\lambda_i)$ are negligible in front of any $\lambda_a-\lambda_j$. It follows that
\begin{equation}
V_j^+-V_j^-\ld\frac{-2ic}{(\lambda_a-\lambda_j)^2+c^2}\,.
\end{equation}
We also have
\begin{equation}
\sqrt{\frac{\prod_{i\neq j}(\lambda_i-\lambda_j+ic)}{\prod_{i\neq j}(\mu_i-\mu_j+ic)}}\ld i^{N-1}\prod_{j\neq a}\sqrt{(\lambda_j-\lambda_a)^2+c^2}\,,
\end{equation}
and
\begin{equation}
\frac{\prod_{i< j}|\lambda_i-\lambda_j|\prod_{i< j}|\mu_i-\mu_j|}{\prod_{i, j}(\mu_j-\lambda_i)}\ld (-1)^{(N-1)(N-2)/2}\prod_{j\neq a}\sign(\lambda_j-\lambda_a)\frac{1}{\prod_{j\neq a}(\mu_j-\lambda_j)}\,.
\end{equation}
As for the matrix $U$, setting $\lambda_p=\lambda_s=\lambda_a$, the dominant entries are
\begin{equation}
U_{aj}\ld-1+\frac{2}{c}\,,
\end{equation}
while the other entries are of order ${\cal O}(L^{-1})$. Hence
\begin{equation}
\underset{i,j}{\det}(\delta_{ij}+U_{ij})\ld\frac{2}{c}\,.
\end{equation}
We obtain the following low-density limit of the form factor
\begin{equation}\label{ldff}
\frac { \left\langle \pmb{\mu} |\psi \left( 0\right) |\pmb{\lambda}\right\rangle  } {\sqrt{\left\langle \pmb{\lambda} \left| \pmb{\lambda} \right\rangle \left\langle \pmb{\mu}\right| \pmb{\mu}\right\rangle }}\ld
\frac{\phi}{\sqrt{L}}\prod_{j\neq a}\frac{2c}{\sqrt{(\lambda_j-\lambda_a)^2+c^2}}\frac{1}{L(\mu_j-\lambda_j)}\,,
\end{equation}
with the phase
\begin{equation}
\phi=(-i)^N \prod_{j\neq a}\sign(\lambda_j-\lambda_a)\,.
\end{equation}

Let us reformulate the meaning of this approximation. We consider $\pmb{\lambda}$ and $\pmb{\mu}$ two Bethe states with $N$ and $N-1$ particles respectively, and denote $\iota:\{1,...,N-1\}\to \{1,...,N\}$ the function such that the element of $\{\lambda_1,...,\lambda_N\}$ that is the closest to $\mu_i$ is $\lambda_{\iota(i)}$. For a dilute $\pmb{\lambda}$, the form factor $\frac { \left\langle \pmb{\mu} |\psi \left( 0\right) |\pmb{\lambda}\right\rangle  } {\sqrt{\left\langle \pmb{\lambda} \left| \pmb{\lambda} \right\rangle \left\langle \pmb{\mu}\right| \pmb{\mu}\right\rangle }}$ is non-negligible in the thermodynamic limit only if $\iota$ is one-to-one from $\{1,...,N-1\}$ to $\{1,...,N\}-\{a\}$ for some $a=1,...,N$. In this case, the form factor reads
\begin{equation}\label{ldff2}
\frac { \left\langle \pmb{\mu} |\psi \left( 0\right) |\pmb{\lambda}\right\rangle  } {\sqrt{\left\langle \pmb{\lambda} \left| \pmb{\lambda} \right\rangle \left\langle \pmb{\mu}\right| \pmb{\mu}\right\rangle }}\ld
\frac{\phi}{\sqrt{L}}\prod_{j=1}^{N-1}\frac{2c}{\sqrt{(\lambda_{\iota(j)}-\lambda_a)^2+c^2}}\frac{1}{L(\mu_j-\lambda_{\iota(j)})}\,.
\end{equation}
Equation \eqref{ldff} corresponds to relabelling $\mu_{\iota^{-1}(j)}$ into $\mu_j$ for $j=1,...,N,\, j\neq a$.\\


Such an expression implies an ordering of the roots $\mu_1,...,\mu_{N-1}$ according to the ordering of the $\lambda_i$'s. Hence in the spectral sum \eqref{bigsumfield}, there is no factor $1/(N-1)!$ once expressed in terms of the Bethe numbers of the $\mu_j$'s. Using this expression for the form factor, one indeed recovers the low-density correlation function \eqref{psipsild2} properly defined  from the partial fraction decomposition, with  ${\cal N}_{\pmb{\lambda}}, {\cal N}_{\pmb{\mu}}=1+{\cal O}(\cal D)$ already imposed at leading order in ${\cal D}$.

\section {Field two-point function\label{fieldsection}}
In this section we compute $S_{0,0,0}$ in \eqref{S00} in the low-density limit.

\subsection{States contributing to the thermodynamic limit}
In order to carry out the sum over the Bethe numbers in \eqref{S00}, let us first investigate which values of Bethe numbers give a non-zero contribution in the thermodynamic limit.

Let us consider Bethe number configurations such that $\mu_i-\lambda_i=\mathcal{O}(L^{-b_i})$ for $i\neq a$ with $ b_i\geq 0$. Since the parity of the number of roots $\lambda_i$ and $\mu_i$ is different, the Bethe numbers of $\lambda_i$ are integers (resp. half-odd integers) if those of $\mu_i$ are half-odd integers (resp. integers). Hence from the Bethe equations it follows that one has $b_i\leq 1$.

Now, since the Bethe number $J_i$ can take $\mathcal{O}(L^{1-b_i})$ values (for $\mu_i-\lambda_i=\mathcal{O}(L^{-b_i})$ to be satisfied), and since $a$ in \eqref{S00} can take $\mathcal{O}(L)$ values, one has $\mathcal{O}(L^{N-\sum_{i\neq a}b_i})$ many such configurations. Besides, each summand in \eqref{S00} is $\mathcal{O}(L^{-2N+1+2\sum_{i\neq a}b_i})$. Hence the contribution of these configurations is $\mathcal{O}(L^{-N+1+\sum_{i\neq a}b_i})$. Given that $0\leq b_i\leq 1$, the only possibility to have a non-vanishing result in the thermodynamic limit is to have $\forall i,\, b_i=1$. Hence the only non-vanishing configurations in \eqref{S00} are those for which the Bethe numbers of $\mu_i$ differ from that of $\lambda_i$ by $\mathcal{O}(L^{0})$. We will denote
\begin{equation}
n_i+\frac{1}{2}=J_i-I_i\qquad \text{for }i\neq a\,,
\end{equation}
the difference between the Bethe numbers of $\mu_i$ and $\lambda_i$, with $n_i$ an integer of order ${\cal O}(L^0)$.

\subsection{Decoupling of the spectral sum in the low-density limit}
The Bethe roots $\mu_i$ involved in \eqref{S00} depend on the difference of Bethe numbers $n_i$ and are all coupled through the Bethe equations. Taking the difference of the Bethe equations \eqref{belog} for $\mu_k$ and $\lambda_k$, one obtains
\begin{equation}
\mu_k-\lambda_k=\frac{2\pi}{L}(G^{-1}\tilde{n})_k+{\cal O}(L^{-2})\,,
\end{equation}
with $G=G(\pmb{\mu})$ introduced in \eqref{gaudin}, and the vector
\begin{equation}
\begin{aligned}
\tilde{n}_k&=n_k+\frac{1}{2}+\frac{1}{\pi}\arctan \frac{\lambda_k-\lambda_a}{c}\,.
\end{aligned}
\end{equation}
In the low-density limit, one has $Gx \ld x$ for all vectors $x$, so that the roots decouple and are expressed as
\begin{equation}
\mu_k-\lambda_k\ld \frac{2\pi}{L}(n_k+\alpha_k(\lambda_a))+{\cal O}(L^{-2})\,,
\end{equation}
where we introduced
\begin{equation}\label{alpha}
\alpha_i(\nu)=\frac{1}{2}+\frac{1}{\pi}\arctan \frac{\lambda_i-\nu}{c}\,.
\end{equation}
Finally, at leading order in ${\cal D}$, the determinants ${\cal N}_{\pmb{\lambda}},{\cal N}_{\pmb{\mu}}$ are equal to $1$ according to \eqref{det1det1}. The spectral sum \eqref{S00} can thus be expressed as a product of $N$ one-dimensional sums
\begin{equation}\label{S001}
\begin{aligned}
S_{0,0,0}\ld\frac{1}{L}&\sum_{a=1}^Ne^{it \lambda_a^2-ix\lambda_a}\\
&\times\prod_{j\neq a}\left(\frac{1}{\pi^2}\frac{c^2}{(\lambda_j-\lambda_a)^2+c^2}\sum_{\substack{n=-\infty}}^{\infty}\frac{e^{i\tfrac{2\pi}{L} (n+\alpha_{j}(\lambda_a))(x-2\lambda_j t)-it\left(\tfrac{2\pi}{L}\right)^2(n+\alpha_j(\lambda_a))^2}}{(n+\alpha_j(\lambda_a))^2}\right)\,.
\end{aligned}
\end{equation}

\subsection{Thermodynamic limit of the spectral sum}
In order to proceed we need to determine the thermodynamic limit of each of the one-dimensional sums, that are of the type
\begin{equation}
\sum_{n\in\mathbb{Z}}\frac{e^{i\frac{w}{L}(n+\alpha)+i\frac{\tau}{L^2}(n+\alpha)^2}}{(n+\alpha)^2}\,,
\end{equation}
for $\alpha,w,\tau$ reals, $\alpha$ non integer. Let us first consider the case $\tau=0$. The quantity $\sum_{n\in\mathbb{Z}}\frac{e^{iWn}}{(n+\alpha)^2}$ is exactly the Fourier series of a certain $2\pi$-periodic function of $W$. Noticing that for $n$ integer
\begin{equation}
\int_{-\pi}^{\pi}\left[\left(\frac{\pi}{\sin \pi\alpha}\right)^2e^{-iW\alpha} +\frac{i\pi}{\sin\pi\alpha}W e^{i\pi\alpha\sign(W)-iW\alpha} \right]e^{-iWn}\D{W}=\frac{2\pi}{(n+\alpha)^2}\,,
\end{equation}
we conclude that
\begin{equation}
\sum_{n\in\mathbb{Z}}\frac{e^{iW(n+\alpha)}}{(n+\alpha)^2}=\left(\frac{\pi}{\sin \pi\alpha}\right)^2 +\frac{i\pi}{\sin\pi\alpha}W e^{i\pi\alpha\sign(W)}\,,\qquad\text{for }-\pi<W\leq\pi\,.
\end{equation}
To treat the case $\tau\neq 0$, we write
\begin{equation}
\sum_{n\in\mathbb{Z}}\frac{e^{i\frac{w}{L}(n+\alpha)+i\frac{\tau}{L^2}(n+\alpha)^2}}{(n+\alpha)^2}=\sum_{n\in\mathbb{Z}}\frac{e^{i\frac{w}{L}(n+\alpha)}}{(n+\alpha)^2}+\frac{1}{L^2}\sum_{n\in\mathbb{Z}}e^{i\frac{w}{L}(n+\alpha)}\frac{e^{i\tau\frac{(n+\alpha)^2}{L^2}}-1}{\tfrac{(n+\alpha)^2}{L^2}}\,.
\end{equation}
The first term was computed previously, and the second one is $1/L$ times the Riemann sum of a regular function, hence in the thermodynamic limit converges to its integral. This yields
\begin{equation}\label{s2s2}
\sum_{n\in\mathbb{Z}}\frac{e^{i\frac{w}{L}(n+\alpha)+i\frac{\tau}{L^2}(n+\alpha)^2}}{(n+\alpha)^2}=\left(\frac{\pi}{\sin \pi\alpha}\right)^2 +\frac{i\pi w}{L\tan(\pi\alpha)}-\frac{\pi|w|}{L}+\frac{\sqrt{|\tau|}}{L}\chi_\pm\left(\tfrac{w}{\sqrt{|\tau|}}\right)+{\cal O}(L^{-2})\,,
\end{equation}
with $\pm=\sign(\tau)$ and
\begin{equation}\label{chi}
\chi_\pm(x)=\int_{-\infty}^\infty e^{ixu}\frac{e^{\pm i u^2}-1}{u^2}\D{u}\,.
\end{equation}
Hence we obtain with $\pm=\sign(t)$, using the expression \eqref{alpha} and trigonometric relations
\begin{equation}
\begin{aligned}
&\sum_{\substack{-\infty<n<\infty}}\frac{e^{\frac{2i\pi}{L} (n+\alpha_{j}(\lambda_a))(x-2\lambda_j t)-\frac{i(2\pi)^2t}{L^2}(n+\alpha_j(\lambda_a))^2}}{(n+\alpha_j(\lambda_a))^2}=\pi^2\frac{(\lambda_j-\lambda_a)^2+c^2}{c^2}\\
&\qquad\times \Bigg[1-\frac{1}{L}\frac {2c^{2}} {\left( \lambda _{j}-\lambda _{a}\right) ^{2}+c^{2}}|x-2\lambda_jt|+\frac{1}{L}\frac {2c^{2}} {\left( \lambda _{j}-\lambda _{a}\right) ^{2}+c^{2}}\frac{\sqrt{|t|}}{\pi}\chi_\mp(\tfrac{x-2\lambda_jt}{\sqrt{|t|}})\\
&\qquad\qquad\qquad\qquad+\frac {2ic(\lambda _{j}-\lambda _{a})} {\left( \lambda _{j}-\lambda _{a}\right) ^{2}+c^{2}}\frac{2\lambda _{j}t-x}{L} \Bigg]+{\cal O}(L^{-2})\,.
\end{aligned}
\end{equation}
Plugging this expression in \eqref{S001}, we use
\begin{equation}\label{expint}
\underset{L\to\infty}{\lim}\, \prod_{i=1}^L \left(1+\frac{f(i/L)}{L}\right)=\exp\left( \int_0^1 f(x)\D{x}\right)\,,
\end{equation}
to obtain in the thermodynamic limit
\begin{equation}\label{field}
\begin {aligned}
 \left\langle \psi ^{\dagger}\left( x,t\right) \psi \left( 0,0\right) \right\rangle\ld\int _{-\infty }^\infty &e^{-it\lambda ^{2}+ix\lambda } \rho \left( \lambda \right)\exp \left (i\int_{-\infty}^\infty \frac {2c(\nu -\lambda )} {\left( \nu -\lambda \right) ^{2}+c^{2}}(2t\nu -x) \rho \left( \nu  \right) \D{\nu}\right)  \\
 &\exp \left (-\int_{-\infty}^\infty \frac {2c^{2}} {\left( \nu -\lambda \right) ^{2}+c^{2}}|x-2\nu t|\rho \left( \nu  \right) \D{\nu}\right) \\
 &\exp \left (\frac{\sqrt{|t|}}{\pi}\int_{-\infty}^\infty \frac {2c^{2}} {\left( \nu -\lambda \right) ^{2}+c^{2}}\chi_\mp(\tfrac{x-2\nu t}{\sqrt{|t|}})\rho \left( \nu  \right) \D{\nu}\right)\D{\lambda} 
  \end {aligned}
\end{equation}
This expression is valid for all space $x$ and time $t$, and all interaction strength $c>0$. It is the leading term in the expansion \eqref{expdensity}, and constitutes the low-density limit of the dynamical correlations of the field.

\subsection {Numerical evaluation \label{numer}}
\begin{figure}[H]
\begin{center}
\begin{tikzpicture}[scale=0.9]
\begin{axis}[
    enlargelimits=false,
    xlabel = $t$,
    xmax=30,
     y tick label style={
        /pgf/number format/.cd,
            fixed,
            fixed zerofill,
            precision=1,
        /tikz/.cd
    }
]
\addplot[
    line width=1pt,
    color=blue,
    opacity=1]
table[
           x expr=\thisrowno{0}, 
           y expr=\thisrowno{1},
         ]{lowdensity_data_c0v25_x0.dat};
\addplot[
    line width=1pt,
    color=blue,
    opacity=0.75]
table[
           x expr=\thisrowno{0}, 
           y expr=\thisrowno{1},
         ]{lowdensity_data_c1_x0.dat};
\addplot[
    line width=1pt,
    color=blue,
    opacity=0.5]
table[
           x expr=\thisrowno{0}, 
           y expr=\thisrowno{1},
         ]{lowdensity_data_c5_x0.dat};
\addplot[
    line width=1pt,
    color=blue,
    opacity=0.25]
table[
           x expr=\thisrowno{0}, 
           y expr=\thisrowno{1},
         ]{lowdensity_data_cinf_x0.dat};
\end{axis}
\end{tikzpicture}
\begin{tikzpicture}[scale=0.9]
\begin{axis}[
    enlargelimits=false,
    xlabel = $t$,
    xmax=30,
     y tick label style={
        /pgf/number format/.cd,
            fixed,
            fixed zerofill,
            precision=1,
        /tikz/.cd
    }
]
\addplot[
    line width=1pt,
    color=blue,
    opacity=0.25]
table[
           x expr=\thisrowno{0}, 
           y expr=\thisrowno{1},
         ]{lowdensity_data_cinf_x2.dat};
\addplot[
    line width=1pt,
    color=blue,
    opacity=0.5]
table[
           x expr=\thisrowno{0}, 
           y expr=\thisrowno{1},
         ]{lowdensity_data_c5_x2.dat};
\addplot[
    line width=1pt,
    color=blue,
    opacity=0.75]
table[
           x expr=\thisrowno{0}, 
           y expr=\thisrowno{1},
         ]{lowdensity_data_c1_x2.dat};
\addplot[
    line width=1pt,
    color=blue,
    opacity=1]
table[
           x expr=\thisrowno{0}, 
           y expr=\thisrowno{1},
         ]{lowdensity_data_c0v25_x2.dat};
\end{axis}
\end{tikzpicture}\\
\caption{Real part of the field two-point function $\Re \left\langle \psi ^{\dagger}\left( x,t\right) \psi \left( 0,0\right) \right\rangle$ in \eqref{field} as a function of $t$, for $x=0$ (left) and $x=2$ (right), and for $c=\infty, 5, 1, \tfrac{1}{4}$ (from light to dark blue). The root density used is given in Section \ref{numer}.}  
\label {fieldfigure}
\end{center}
\end {figure}
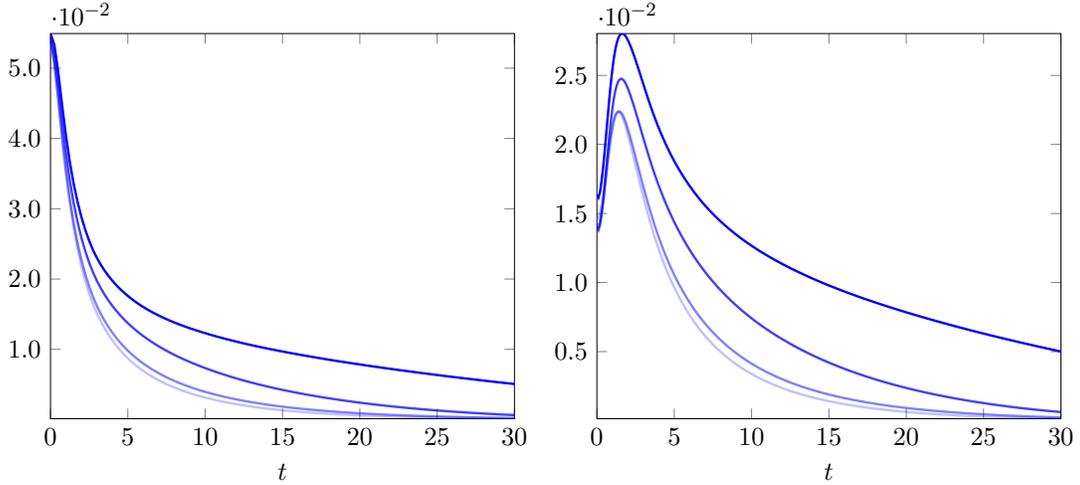

In this section we evaluate the correlation function \eqref{field} for a root density corresponding to a thermal state\cite{YangYang69}.  These are parametrized by the inverse temperature $\beta$ and the particle density ${\cal D}$. Defining the so-called dressed energy 
$\varepsilon_{\rm dr}(\lambda)$ by 
\begin{equation}
\frac{\rho(\lambda)}{\rho(\lambda)+\rho_h(\lambda)}=\frac{1}{1+e^{\beta\varepsilon_{\rm dr}(\lambda)}}\,,
\label{dresseden}
\end{equation}
the filling function $n(\lambda)$ of a thermal state is such that
\begin{equation}
\varepsilon_{\rm dr}(\lambda)=\lambda^2-h-\frac{1}{2\pi\beta}\int_{-\infty}^\infty \frac{2c}{c^2+(\lambda-\mu)^2}\log(1+e^{-\beta\varepsilon_{\rm dr}(\mu)})\D{\mu}\,.
\label{dresseden2}
\end{equation}
Here $h$ is a chemical potential that is used to fix the desired particle density ${\cal D}$. In Figure \ref{fieldfigure} we evaluate \eqref{field} as a function of $t$ at fixed $x$, for several values of $c$, and for a thermal state with ${\cal D}=0.055$ and $\beta=1$.

\section {Density two-point function\label{densitysection}}
In this Section we apply the same reasoning as in Sections \ref{ldd} and \ref{fieldsection} but to the density two-point function.

\subsection{Partial fraction decomposition\label{partyop12}}
We consider $ |\pmb{\lambda} \rangle $ and $ |\pmb{\mu} \rangle $ eigenstates of the Lieb-Liniger Hamiltonian with $N$ rapidities, and define $F_\sigma(\pmb{\lambda},\pmb{\mu})$ by
\begin{equation}
\begin{aligned}
\frac {\left| \left\langle \pmb{\mu} |\sigma \left( 0\right) |\pmb{\lambda}\right\rangle \right| ^{2}} {\left\langle \pmb{\lambda} \left| \pmb{\lambda} \right\rangle \left\langle \pmb{\mu}\right| \pmb{\mu}\right\rangle }= \frac{F_\sigma(\pmb{\lambda},\pmb{\mu})}{L^{2N}\mathcal{N}_{\pmb{\lambda}}\mathcal{N}_{\pmb{\mu}}} \,.
\end{aligned}
\end{equation}
Similarly to the field case $F_\psi$, the reduced form factor $F_\sigma(\pmb{\lambda},\pmb{\mu})$ is a ratio of two polynomials in the Bethe roots so that it can be decomposed into partial fractions. Repeating the arguments that apply to the field case, we similarly obtain the writing
\begin{equation}\label{pfdsigma}
F_\sigma(\pmb{\lambda},\pmb{\mu})=\sum_{\{\nu\},f}\frac{A(\pmb{\lambda},\pmb{\mu},\{\nu\},f)}{\prod_{i=1}^{N-1}(\mu_i-\lambda_{f(i)})^{\nu_i}}\,,
\end{equation}
where each $\nu_i$ takes the value $0,1$ or $2$, and where $f$ are functions $\{i\in\{1,...,N\}| \nu_i\neq 0\}\to \{1,...,N\}$. The coefficients $A(\pmb{\lambda},\pmb{\mu},\{\nu\},f)$ crucially do not depend on any $\mu_i$ whenever $\nu_i>0$, and are bounded functions of real $\mu_i$ if $\nu_i=0$. The function $f$ has the same constraints as in the field case, namely it can take at most twice the same value, and takes only once a value on points with $\nu_i=2$.

This decomposition also leads to an expansion in the particle density ${\cal D}$. Namely, with the parametrization $m=|I_1|$, $n=|I_2|$, $p=|\{i| \nu_i=0\}|$
\begin{equation}
\begin{aligned}
I_2&=\{j_{1},...,j_{n}\}\\
I_0&=\{j_{n+1},...,j_{2n+p}\}\\
I_1&=\{j_{2n+p+1},...,j_{2n+p+m}\}\\
\{1,...,N\}-(I_0\cup I_1 \cup I_2)&=\{j_{2n+p+m+1},...,j_{N}\}\,.
\end{aligned}
\end{equation}
we have
\begin{equation}\label{expdensitysigma}
\left\langle  \sigma\left( x,t\right) \sigma \left( 0,0\right)  \right\rangle =\sum_{n,m,p\geq 0}S_{n,m,p}\,,
\end{equation}
with
\begin{equation}
\begin{aligned}
&S_{n,m,p}=\frac{1}{2^np!L^{2N}}\sum_{\substack{I_{0,1,2}\subset \{1,...,N\}\\|I_0|=n+p\\|I_1|=m\\|I_2|=n \\\text{all disjoint}}}\sum_{J_1,...,J_{N}}\frac{{\cal A}(I_0,I_1,I_2|\{\mu_i\}_{i=2n+1}^{2n+p})}{\mathcal{N}_{\pmb{\lambda}}\mathcal{N}_{\pmb{\mu}}}\\
  &\qquad\qquad\times\frac{e^{it\left( E\left( \pmb{\lambda}\right) -E\left( \pmb{\mu} \right) \right) +ix\left( P\left( \pmb{\mu}\right) -P\left( \pmb{\lambda}\right) \right) }}{\prod_{i=1}^{n}(\mu_{2i-1}-\lambda_{j_i})(\mu_{2i}-\lambda_{j_i})\prod_{i=2n+1+p}^{2n+m+p}(\mu_i-\lambda_{j_{i}})\prod_{i=2n+m+p+1}^{N}(\mu_i-\lambda_{j_{i}})^{2}}\,.
\end{aligned}
\end{equation}
 The leading term is a priori obtained with $\nu_i=2$, and the next terms by replacing some of the $\nu_i$'s by $1$ or $0$, each time picking a factor ${\cal D}$. However, in contrast to the field case, the coefficient of the a priori leading term is actually found to vanish, using the analogue of \eqref{formulA} for the density operator case
 \begin{equation}
{\cal A}(\varnothing,\varnothing,\varnothing|\varnothing)=0\,.
\end{equation}
If one $\nu_i$ is set to $1$, the coefficient still vanishes
 \begin{equation}
{\cal A}(\varnothing,\{a\},\varnothing|\varnothing)=0\,.
\end{equation}
However, if it is set to $0$, the coefficient is non-zero and reads
 \begin{equation}
{\cal A}(\{a\},\varnothing,\varnothing|\mu_a)=\prod_{j\neq a}\frac{4c^2(\lambda_a-\mu_a)^2}{[(\lambda_j-\lambda_a)^2+c^2][(\lambda_j-\mu_a)^2+c^2]}\,.
\end{equation}
Hence the leading order in the density expansion, i.e. the low-density limit, is obtained with
\begin{equation}
\left\langle  \sigma( x,t) \sigma( 0,0)  \right\rangle\ld S_{0,0,1}\,.
\end{equation}
It yields the following low-density limit of the density two-point function
\begin{equation}\label{2pfdensityld}
\begin{aligned}
 & \left\langle  \sigma( x,t) \sigma( 0,0)  \right\rangle \ld \frac{1}{L^2}\sum_{\lambda_a,\mu_a}e^{ix(\mu_a-\lambda_a)+it(\lambda_a^2-\mu_a^2)}\\&\times\sum _{J_i,\, i\neq a}\frac{1}{\mathcal{N}_{\pmb{\lambda}}\mathcal{N}_{\pmb{\mu}}}\prod_{j\neq a}\frac{c^2}{(\lambda_j-\lambda_a)^2+c^2}\frac{c^2}{(\lambda_j-\mu_a)^2+c^2}\cdot
  \frac{4(\lambda_a-\mu_a)^2}{c^2L^2(\mu_j-\lambda_j)^2}e^{ix(\mu_j-\lambda_j)+it(\lambda_j^2-\mu_j^2)}\,.
\end{aligned}
\end{equation}

\subsection {Interpretation: low-density limit of the density form factor}
Similarly to the field operator, the low-density limit of the density two-point function is defined in terms of a partial fraction decomposition of the form factor. In the field case, one was also able to obtain a low-density limit of the field form factor by considering dilute states, see Section \ref{intuitive}. In the density case, we are going to follow a different route in order to show the usefulness of \eqref{ldff}, by recovering the low-density limit of the density correlation \eqref{2pfdensityld} from the low-density approximation of the field form factor  \eqref{ldff}.
%
 The rationale is that  the density form factor can be obtained as a limit of a field two-point function between different eigenstates $\pmb{\lambda}$ and $\pmb{\mu}$ with $N$ roots
\begin{equation}
\frac{\langle \pmb{\mu}|\sigma(0)|\pmb{\lambda}\rangle}{\sqrt{\langle \pmb{\mu}|\pmb{\mu}\rangle \langle \pmb{\lambda}|\pmb{\lambda}\rangle}}=\underset{x\to 0}{\lim}\, \frac{\langle \pmb{\mu}|\psi^\dagger( x,0) \psi ( 0,0)|\pmb{\lambda}\rangle}{\sqrt{\langle \pmb{\mu}|\pmb{\mu}\rangle \langle \pmb{\lambda}|\pmb{\lambda}\rangle}}\,.
\end{equation}
This two-point function can itself be expressed as a Lehmann representation
\begin{equation}
\frac{\langle \pmb{\mu}|\psi^\dagger( x,0) \psi ( 0,0)|\pmb{\lambda}\rangle}{\sqrt{\langle \pmb{\mu}|\pmb{\mu}\rangle \langle \pmb{\lambda}|\pmb{\lambda}\rangle}}=\sum_{\pmb{\nu}}\frac{\langle \pmb{\nu}|\psi(0)|\pmb{\mu}\rangle^* \langle \pmb{\nu}|\psi(0)|\pmb{\lambda}\rangle}{\sqrt{\langle \pmb{\mu}|\pmb{\mu}\rangle \langle \pmb{\lambda}|\pmb{\lambda}\rangle}\langle \pmb{\nu}|\pmb{\nu}\rangle}e^{ix(P(\pmb{\nu})-P(\pmb{\mu}))}\,,
\end{equation}
where $\pmb{\nu}$ is a state with $N-1$ roots. In the low-density limit, according to \eqref{ldff}, for the field form factors not to vanish one has to have $N-1$ roots $\lambda_i$ with exactly one $\nu_j$ around at a distance ${\cal O}(L^{-1})$, leaving a $\lambda_a$ without $\nu_j$'s around. The same holds for the $\mu_i$'s with respect to the $\nu_j$'s. This implies that in the low-density limit, there is also exactly one $\mu_i$ around each $\lambda_j$ at a distance ${\cal O}(L^{-1})$ for $j\neq a$, leaving a $\mu_a$ without $\nu_j$'s nor $\lambda_j$'s around. Since $\pmb{\mu}$ and $\pmb{\lambda}$ are an input of the problem, these $\mu_a$ and $\lambda_a$ are fixed by the choices of $\pmb{\lambda}$ and $\pmb{\mu}$ (they are not free parameters like in the two-point function case with $\pmb{\lambda}=\pmb{\mu}$). Then using  \eqref{ldff} we obtain the following low-density limit
\begin{equation}
\begin{aligned}
&\frac{\langle \pmb{\mu}|\psi^\dagger( x,0) \psi ( 0,0)|\pmb{\lambda}\rangle}{\sqrt{\langle \pmb{\mu}|\pmb{\mu}\rangle \langle \pmb{\lambda}|\pmb{\lambda}\rangle}}\ld \\
&\qquad\qquad\frac{e^{-ix\mu_a}}{L}\sum_{\pmb{\nu}}\prod_{j\neq a}\frac{2c}{\sqrt{(\lambda_j-\lambda_a)^2+c^2}}\frac{2c}{\sqrt{(\mu_j-\mu_a)^2+c^2}} \frac{1}{L(\nu_j-\lambda_j)}\frac{e^{ix(\nu_j-\mu_j)}}{L(\nu_j-\mu_j)}\,.
\end{aligned}
\end{equation}
The Bethe equations allow us to write
\begin{equation}
\begin{aligned}
\nu_j-\lambda_j&\ld \frac{2\pi}{L}(n_j+\alpha_j(\lambda_a))\\
\nu_j-\mu_j&\ld \frac{2\pi}{L}(n_j+\alpha_j(\mu_a)-p_j)\,,
\end{aligned}
\end{equation}
with $n_j,p_j$ integers. The integer $p_j$ is related to $\mu_j$ and $\lambda_j$ through
\begin{equation}
\mu_j-\lambda_j\ld \frac{2\pi}{L}(p_j+\alpha_j(\lambda_a)-\alpha_j(\mu_a))\,,
\end{equation}
which is a parameter of the problem that is not to be summed over. We obtain the following factorization
\begin{equation}
\begin{aligned}
\frac{\langle \pmb{\mu}|\psi^\dagger( x,0) \psi ( 0,0)|\pmb{\lambda}\rangle}{\sqrt{\langle \pmb{\mu}|\pmb{\mu}\rangle \langle \pmb{\lambda}|\pmb{\lambda}\rangle}}\ld \frac{e^{-ix\mu_a}}{L}&\prod_{j\neq a}\frac{c^2}{\pi^2}\frac{1}{\sqrt{(\lambda_j-\lambda_a)^2+c^2}}\frac{1}{\sqrt{(\lambda_j-\mu_a+{\cal O}(L^{-1}))^2+c^2}} \\
&\times\sum_{n=-\infty}^\infty\frac{e^{\frac{2i \pi x}{L}(n+\alpha_j(\mu_a)-p_j)}}{(n+\alpha_j(\lambda_a))(n+\alpha_j(\mu_a)-p_j)}\,.
\end{aligned}
\end{equation}
In order to determine the thermodynamic limit of the field four-point function we need to carry out the sums over the integers, that reduces to sums of the type
\begin{equation}
\sum_{n\in\mathbb{Z}}\frac{e^{i\tfrac{w}{L}(n+\alpha)}}{n+\alpha}\,.
\end{equation}
We notice that $\sum_{n\in\mathbb{Z}}\frac{e^{iWn}}{n+\alpha}$ is the Fourier series of a $2\pi$-periodic function of $W$, and
\begin{equation}
\int_{-\pi}^\pi \frac{\pi}{\sin \pi\alpha}e^{i\pi\alpha\sign(W)-iW\alpha} e^{-iWn}\D{W}=\frac{2\pi}{n+\alpha}\,.
\end{equation}
Hence
\begin{equation}
\sum_{n\in\mathbb{Z}}\frac{e^{iW(n+\alpha)}}{n+\alpha}=\frac{\pi}{\sin \pi\alpha}e^{i\pi\alpha\sign(W)}\,,\qquad\text{for }-\pi<W\leq\pi\,.
\end{equation}
To use this formula we write
\begin{equation}
\frac{1}{(n+\alpha_j(\lambda_a))(n+\alpha_j(\mu_a)-p_j)}=\frac{1}{p_j+\alpha_j(\lambda_a)-\alpha_j(\mu_a)}\left(\frac{1}{n+\alpha_j(\mu_a)-p_j} -\frac{1}{n+\alpha_j(\lambda_a)}\right)\,.
\end{equation}
It yields
\begin{equation}
\begin{aligned}
\frac{\langle \pmb{\mu}|\psi^\dagger( x,0) \psi ( 0,0)|\pmb{\lambda}\rangle}{\sqrt{\langle \pmb{\mu}|\pmb{\mu}\rangle \langle \pmb{\lambda}|\pmb{\lambda}\rangle}}\ld &\frac{e^{-ix\mu_a}}{L}\prod_{j\neq a}\frac{c^2}{\pi}\frac{1}{\sqrt{(\lambda_j-\lambda_a)^2+c^2}}\frac{1}{\sqrt{(\lambda_j-\mu_a)^2+c^2}} \\
&\times\frac{1}{p+\alpha_j(\lambda_a)-\alpha_j(\mu_a)}\Bigg[\frac{1}{\tan \pi\alpha_j(\mu_a)}-\frac{e^{\frac{2i\pi x}{L}(\alpha_j(\mu_a)-\alpha_j(\lambda_a)-p_j)}}{\tan \pi\alpha_j(\lambda_a)} \\
&\qquad+i\sign(x) \left( 1-e^{\frac{2i\pi x}{L}(\alpha_j(\mu_a)-\alpha_j(\lambda_a)-p_j)}\right)\Bigg]\,.
\end{aligned}
\end{equation}
We observe that the limit $x\to 0$ is regular. Using the expression for $\alpha_j(\nu)$ \eqref{alpha}, we obtain the following low-density limit of the form factor
\begin{equation}\label{lddensity}
\begin{aligned}
\frac{\langle \pmb{\mu}|\sigma(0)|\pmb{\lambda}\rangle}{\sqrt{\langle \pmb{\mu}|\pmb{\mu}\rangle \langle \pmb{\lambda}|\pmb{\lambda}\rangle}}\ld \frac{1}{L}&\prod_{j\neq a}\frac{c}{\sqrt{(\lambda_j-\lambda_a)^2+c^2}}\frac{c}{\sqrt{(\lambda_j-\mu_a)^2+c^2}}\frac{2(\mu_a-\lambda_a)}{cL(\mu_j-\lambda_j)}\,.
\end{aligned}
\end{equation}
Plugging this expression into the Lehmann representation for the density two-point function \eqref{bigsumfield}, one obtains exactly the low-density limit \eqref{2pfdensityld} previously obtained from the partial fraction decomposition. This provides an interesting consistency check of the interpretations of the low-density limit as obtained from low-density approximations of form factors.

\subsection {The thermodynamic limit of the spectral sum}
We now wish to compute the spectral sum \eqref{2pfdensityld} in the low-density limit. Repeating the arguments performed for the field case, in the thermodynamic limit only the states with $\mu_j-\lambda_j={\cal O}(L^{-1})$ contribute in \eqref{2pfdensityld}. Besides, like for the field correlations, the Bethe equations decouple in the low-density limit. By taking the difference of the Bethe equations, we have for $j\neq a$
\begin{equation}
\mu_j-\lambda_j\ld \frac{2\pi}{L}(p_j+\alpha_j(\lambda_a)-\alpha_j(\mu_a))\,,
\end{equation}
with $p_j$ an integer. Using ${\cal N}_{\pmb{\lambda}},{\cal N}_{\pmb{\mu}}\ld 1$, we obtain the following factorization of the spectral sum
\begin{equation}\label{specsum}
\begin{aligned}
  \left\langle  \sigma( x,t) \sigma( 0,0)  \right\rangle &\ld \frac{1}{L^2}\sum_{\lambda_a,\mu_a}e^{ix(\mu_a-\lambda_a)+it(\lambda_a^2-\mu_a^2)}\prod_{j\neq a}\frac{c^2}{(\lambda_j-\lambda_a)^2+c^2}\frac{c^2}{(\lambda_j-\mu_a)^2+c^2}\\
  &\frac{(\lambda_a-\mu_a)^2}{\pi^2c^2}\sum_{p=-\infty}^\infty\frac{e^{\frac{2i\pi (x-2\lambda_j t)}{L}(p+\alpha_j(\lambda_a)-\alpha_j(\mu_a))-\frac{it(2\pi)^2}{L^2}(p+\alpha_j(\lambda_a)-\alpha_j(\mu_a))^2}}{(p+\alpha_j(\lambda_a)-\alpha_j(\mu_a))^2}\,.
\end{aligned}
\end{equation}
We now use \eqref{s2s2} to compute the sum over the integers $p$. We find
\begin{equation}
\begin{aligned}
&\sum_{p=-\infty}^\infty\frac{e^{\frac{2i\pi (x-2\lambda_j t)}{L}(p+\alpha_j(\lambda_a)-\alpha_j(\mu_a))-\frac{it(2\pi)^2}{L^2}(p+\alpha_j(\lambda_a)-\alpha_j(\mu_a))^2}}{(p+\alpha_j(\lambda_a)-\alpha_j(\mu_a))^2}=\\
&\frac{\pi^2 c^2 (1+\tfrac{(\lambda_a-\lambda_j)^2}{c^2})(1+\tfrac{(\mu_a-\lambda_j)^2}{c^2})}{(\lambda_a-\mu_a)^2}\Bigg[1+\frac{2i(x-2\lambda_j t)}{Lc} (\mu_a-\lambda_a)\frac{1+\tfrac{(\mu_a-\lambda_j)(\lambda_a-\lambda_j)}{c^2}}{(1+\tfrac{(\lambda_a-\lambda_j)^2}{c^2})(1+\tfrac{(\mu_a-\lambda_j)^2}{c^2})}\\
&\qquad\qquad\qquad-\frac{2|x-2\lambda_j t|}{Lc^2}\frac{(\lambda_a-\mu_a)^2}{(1+\tfrac{(\lambda_a-\lambda_j)^2}{c^2})(1+\tfrac{(\mu_a-\lambda_j)^2}{c^2})}\\
&\qquad\qquad\qquad+\frac{2\sqrt{|t|}}{\pi Lc^2}\frac{(\lambda_a-\mu_a)^2}{(1+\tfrac{(\lambda_a-\lambda_j)^2}{c^2})(1+\tfrac{(\mu_a-\lambda_j)^2}{c^2})} \chi_\mp\left(\tfrac{x-2\lambda_j t}{\sqrt{|t|}}\right)\Bigg]+{\cal O}(L^{-2})
\end{aligned}
\end{equation}
Plugging this relation in \eqref{specsum}, the product over $j$ becomes an exponential in the thermodynamic limit, according to \eqref{expint}. The sum over the roots of the representative state $\lambda_a$ becomes an integral over the root density $\rho(\lambda)$, while the sum over $\nu_a$ that has to differ from any $\lambda_j$ becomes an integral over the hole density $\rho_h(\lambda)$ defined in \eqref{vartheta}. Hence one obtains
\begin{equation}\label{densityde}
\begin{aligned}
 \left\langle  \sigma( x,t) \sigma( 0,0)  \right\rangle& \ld \int_{-\infty}^\infty  \int_{-\infty}^\infty e^{it(\lambda^2-\mu^2)+ix(\mu-\lambda)}\rho(\lambda)\rho_h(\mu)\\
 &\exp \left(2ic(\mu-\lambda)\int_{-\infty}^\infty \frac{(c^2+(\mu-\nu)(\lambda-\nu))}{(c^2+(\lambda-\nu)^2)(c^2+(\mu-\nu)^2)}(x-2\nu t)\rho(\nu)\D{\nu} \right)\\
 &\exp \left(-2c^2(\mu-\lambda)^2\int_{-\infty}^\infty \frac{|x-2\nu t|}{(c^2+(\lambda-\nu)^2)(c^2+(\mu-\nu)^2)} \rho(\nu)\D{\nu} \right)\\
 &\exp \left(\frac{2c^2\sqrt{|t|}}{\pi}(\mu-\lambda)^2\int_{-\infty}^\infty \frac{\chi_\mp(\tfrac{x-2\nu t}{\sqrt{|t|}})}{(c^2+(\lambda-\nu)^2)(c^2+(\mu-\nu)^2)} \rho(\nu)\D{\nu} \right)\D{\lambda} \D{\mu}\,.
\end{aligned}
\end{equation}
This expression is valid for all space $x$ and time $t$, and all interaction strength $c>0$. It is the leading term in the expansion \eqref{expdensitysigma}, and is the low-density limit of the dynamical correlations of the density.

\section{Comments\label{comments}}
\subsection{Bare particle-hole excitations\label{which}}
In the spectral sum \eqref{bigsumfield} the intermediate states $\pmb{\mu}$ can be seen as excited states above the averaging state $\pmb{\lambda}$. In this picture it is natural to \textit{expand} the spectral sum \eqref{bigsumfield} in terms of the number of \textit{particle-hole excitations} that $\pmb{\mu}$ has over $\pmb{\lambda}$. Such an expansion consists in writing
\begin{equation}
\left\langle  \sigma( x,t) \sigma( 0,0)  \right\rangle={\cal D}^2+\sum_{n\geq 1}\mathfrak{S}_n
\end{equation}
with
\begin{equation}\label{bare}
\mathfrak{S}_n=\sum _{ \substack{\pmb{\mu}\\ |\{I_a\}\cap \{J_a\}|=N-n}}\frac {\left| \left\langle \pmb{\mu} |\sigma( 0) |\pmb{\lambda}\right\rangle \right| ^{2}} {\left\langle \pmb{\lambda} \left| \pmb{\lambda} \right\rangle \left\langle \pmb{\mu}\right| \pmb{\mu}\right\rangle }e^{it\left( E\left( \pmb{\lambda}\right) -E\left( \pmb{\mu} \right) \right) +ix\left( P\left( \pmb{\mu}\right) -P\left( \pmb{\lambda}\right) \right) }\,,
\end{equation}
which is the spectral sum restricted to intermediate states $\pmb{\mu}$ that share $N-n$ Bethe numbers $J_a$ with the Bethe numbers $I_a$ of $\pmb{\lambda}$. Ideally, each individual $\mathfrak{S}_n$ would have a well-defined and finite thermodynamic limit $L\to\infty$ that could be represented as a multiple integral over root densities and hole densities
\begin{equation}\label{multip}
\begin{aligned}
\underbrace{\int_{-\infty}^\infty...\int_{-\infty}^\infty}_{2n}&F_n(\lambda_1,\mu_1,...,\lambda_n,\mu_n)e^{it(E(\pmb{\lambda})-E(\pmb{\mu}))+ix(P(\pmb{\mu})-P(\pmb{\lambda}))}\\
&\qquad\qquad\qquad\times\rho(\lambda_1)\rho_h(\mu_1)...\rho(\lambda_n)\rho_h(\mu_n)\D{\lambda_1}\D{\mu_1}...\D{\lambda_n}\D{\mu_n}\,.
\end{aligned}
\end{equation}
The function $F_n(\lambda_1,\mu_1,...,\lambda_n,\mu_n)$ appearing in this expression would thus be identified as a \textit{thermodynamic form factor} for $n$ particle-hole excitations\cite{deNP15,deNP16,DNP18,panfil20}. This idea was backed by calculations of
\begin{equation}
\underset{\mu\to\lambda}{\lim}\,F_1(\lambda,\mu)\,,
\end{equation}
that is finite and well-defined in the thermodynamic limit, by various means \cite{deNP16,cortescuberopanfil2}.

However, this picture seems to be a priori contradicted by results of \cite{granetessler20}, where the thermodynamic limit of the spectral sum \eqref{bigsumfield} was computed exactly at order $c^{-2}$. At this order, the spectral sum \textit{exactly} truncates to one- and two-particle-hole excitations, hence to $\mathfrak{S}_1$ and $\mathfrak{S}_2$. But it was found that these two separate sums in the thermodynamic limit individually \textit{diverge} and \textit{depend on the representative state} $\pmb{\lambda}$ of the root density, at order $c^{-2}$. Namely we have
\begin{equation}
\begin{aligned}\label{s1s2}
&\mathfrak{S}_1=LA_1+f_1(\rho,\gamma_{-2})+{\cal O}(L^{-1})+{\cal O}(c^{-3})\\
&\mathfrak{S}_2=LA_2+f_2(\rho,\gamma_{-2})+{\cal O}(L^{-1})+{\cal O}(c^{-3})\,,
\end{aligned}
\end{equation}
with $A_{1,2}$ some reals of order $c^{-2}$ and $f_{1,2}(\rho,\gamma_{-2})$ functions of the root density and the pair distribution function (see \cite{granetessler20} for a precise definition). However, their sum $\mathfrak{S}_1+\mathfrak{S}_2$ is not divergent and depends only on $\rho$, as we expect. To fit in the general picture \eqref{multip}, the only solution would be that these divergences in $L$ are an artefact of the $1/c$ expansion, i.e. that the finite-size corrections to $\mathfrak{S}_{1,2}$ in the thermodynamic limit would be e.g. of the form $\varphi(L/c)$ with a function $\varphi(x)\to 0$ when $x\to \pm\infty$. In the framework of the low-density expansion, one can investigate this issue since the calculations are performed non-perturbatively in $1/c$, and compute the thermodynamic limit of $\mathfrak{S}_{1}$ for a finite fixed $c$, in the low-density limit.\\

To that end, let us use the expression \eqref{lddensity} for the low-density limit of the form factor to write
\begin{equation}\label{lowdensquare}
\begin{aligned}
\frac{|\langle \pmb{\mu}|\sigma(0)|\pmb{\lambda}\rangle|^2}{\langle \pmb{\mu}|\pmb{\mu}\rangle \langle \pmb{\lambda}|\pmb{\lambda}\rangle}\ld \frac{1}{L^2}&\prod_{j\neq a}\frac{c^2}{(\lambda_j-\lambda_a)^2+c^2}\frac{c^2}{(\lambda_j-\mu_a)^2+c^2}\frac{4(\mu_a-\lambda_a)^2}{c^2L^2(\mu_j-\lambda_j)^2}\,,
\end{aligned}
\end{equation}
and assume that $\pmb{\mu}$ only involves one particle-hole excitation above $\pmb{\lambda}$. Hence for all $j\neq a$, the Bethe numbers $J_j$ of $\pmb{\mu}$ are equal to the Bethe numbers $I_j$ of $\pmb{\lambda}$. In the low-density limit we have then
\begin{equation}
\forall j\neq a,\qquad\mu_j-\lambda_j\ld \frac{2\pi}{L}(\alpha_j(\lambda_a)-\alpha_j(\mu_a))\,.
\end{equation}
It follows that in the thermodynamic limit, the low-density form factor squared \eqref{lowdensquare} becomes
\begin{equation}
\frac{|\langle \pmb{\mu}|\sigma(0)|\pmb{\lambda}\rangle|^2}{\langle \pmb{\mu}|\pmb{\mu}\rangle \langle \pmb{\lambda}|\pmb{\lambda}\rangle}\ld \frac{1}{L^2}\exp[L\phi(\lambda_a,\mu_a)+{\cal O}(L^0)]\,,
\end{equation}
with
\begin{equation}\label{phi}
\phi(\lambda,\mu)=\int_{-\infty}^\infty \log \left[\frac{1}{c^2}\frac{1}{1+\tfrac{(\nu-\lambda)^2}{c^2}} \frac{1}{1+\tfrac{(\nu-\mu)^2}{c^2}}\left( \frac{\mu-\lambda}{\arctan \tfrac{\mu-\nu}{c}-\arctan \tfrac{\lambda-\nu}{c}}\right)^2\right] \rho(\nu)\D{\nu}\,.
\end{equation}
The sign of the logarithm can be determined by applying the following inequality that is valid for all real $u$
\begin{equation}\label{ineq}
u^2\geq \sin^2 u\,,
\end{equation}
to the value
\begin{equation}
u=\arctan x-\arctan y\,,
\end{equation}
for $x,y$ reals. Indeed, using trigonometric relations, Eq \eqref{ineq} is exactly
\begin{equation}
\left( \frac{\arctan x-\arctan y}{x-y}\right)^2\geq \arctan'(x)\arctan'(y)\,,
\end{equation}
with equality only for $x=y$. This permits to deduce that the logarithm in \eqref{phi} is always negative whenever $\lambda\neq \mu$. One concludes that we always have if $\lambda\neq \mu$
\begin{equation}
\phi(\lambda,\mu)<0\,,
\end{equation}
and if $\lambda=\mu$ we have
\begin{equation}
\phi(\lambda,\lambda)=0\,.
\end{equation}
In other words, the bare particle-hole excitations for $\lambda\neq\mu$ are exponentially small in $L$. It follows that in the framework of a multiple integral representation in terms of bare particle-hole excitations in the thermodynamic limit \eqref{multip}, the function $F_1(\lambda,\mu)$  would be in fact zero everywhere except if $\mu=\lambda$ where it takes the value $1$ as deduced from \eqref{lowdensquare}. Namely we would have
\begin{equation}
F_1(\lambda,\mu)= \begin{cases}0\qquad \text{if }\lambda\neq \mu\\
1\qquad \text{if }\lambda= \mu
\end{cases}+{\cal O}({\cal D})\,.
\end{equation}

This analysis shows that the integral representation \eqref{multip} in terms of 'bare' particle-hole excitations \eqref{bare} is singular. If the leading behaviour at large space and time of \eqref{multip} is indeed obtained when $\lambda_1=\mu_1$ in \eqref{multip} for $n=1$, where the function $F_1(\lambda,\mu)$ takes a finite value, the singular behaviour of this representation should be an obstacle to the computation of the subleading orders.

\subsection{Dressed particle-hole excitations\label{which2}}
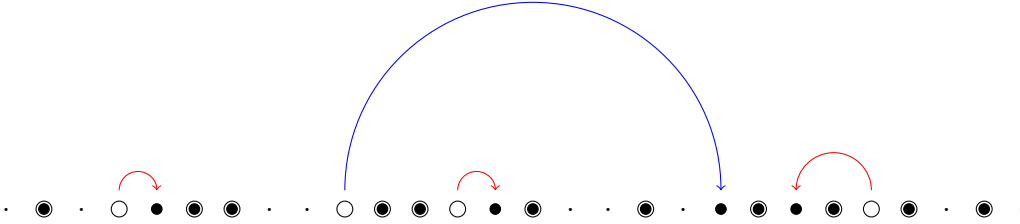
\begin{figure}[H]
\begin{center}
\begin{tikzpicture}[scale=1]
\draw[->,blue]     (3.5,0.25) arc (180: 0:2.5);
\draw[->,red]     (0.5,0.25) arc (180: 0:0.25);
\draw[->,red]     (5.,0.25) arc (180: 0:0.25);
\draw[->,red]     (10.5,0.25) arc (0: 180:0.5);
\node at (-1,0) {.};
\draw[black] (-0.5,0) circle (3pt);
\node at (0,0) {.};
\draw[black] (0.5,0) circle (3pt);
\node at (1,0) {.};
\draw[black] (1.5,0) circle (3pt);
\draw[black] (2,0) circle (3pt);
\node at (2.5,0) {.};
\node at (3,0) {.};
\draw[black] (3.5,0) circle (3pt);
\draw[black] (4,0) circle (3pt);
\draw[black] (4.5,0) circle (3pt);
\draw[black] (5,0) circle (3pt);
\node at (5.5,0) {.};
\draw[black] (6,0) circle (3pt);
\node at (6.5,0) {.};
\node at (7,0) {.};
\draw[black] (7.5,0) circle (3pt);
\node at (8,0) {.};
\node at (8.5,0) {.};
\draw[black] (9,0) circle (3pt);
\node at (9.5,0) {.};
\draw[black] (10,0) circle (3pt);
\draw[black] (10.5,0) circle (3pt);
\draw[black] (11,0) circle (3pt);
\node at (11.5,0) {.};
\draw[black] (12,0) circle (3pt);
\node at (12.5,0) {.};
\filldraw[black] (-0.5,0) circle (2pt);
\filldraw[black] (1.,0) circle (2pt);
\filldraw[black] (1.5,0) circle (2pt);
\filldraw[black] (2,0) circle (2pt);
\filldraw[black] (8.5,0) circle (2pt);
\filldraw[black] (4,0) circle (2pt);
\filldraw[black] (4.5,0) circle (2pt);
\filldraw[black] (5.5,0) circle (2pt);
\filldraw[black] (6,0) circle (2pt);
\filldraw[black] (9,0) circle (2pt);
\filldraw[black] (7.5,0) circle (2pt);
\filldraw[black] (10,0) circle (2pt);
\filldraw[black] (9.5,0) circle (2pt);
\filldraw[black] (11,0) circle (2pt);
\filldraw[black] (12,0) circle (2pt);
\end{tikzpicture}
\end{center}
\caption{Sketch of a dressed one-particle-hole excitation: positions of the
momenta of the averaging state $\pmb{\lambda}$ (empty circles) and the
intermediate state $\pmb{\mu}$  (filled circles) respectively, and position of 
the holes (dots). In red is indicated the 'soft modes' corresponding to microscopic excitations, and in blue the only macroscopic excitation.} 
\label{exfnp1k0}
\end{figure}

Let us now investigate the nature of the states summed over to obtain the expression \eqref{densityde} in the low-density limit. The double integral over the root density and hole density are one-particle-hole excitations with a \textit{macroscopic} amplitude (i.e. the difference between the Bethe numbers is ${\cal O}(L)$). The exponentials however arise from the product of a macroscopic number of one-dimensional sums over all the other remaining roots of the averaging states. These take into account an \textit{arbitrary} number of particle-hole excitations, but with only a \textit{microscopic} or  \textit{mesoscopic} amplitude (i.e. the difference between the Bethe numbers can be ${\cal O}(L^\nu)$ for any $\nu<1$). These configurations are represented in Figure \ref{exfnp1k0}. They correspond to what is called a \textit{dressed one-particle-hole excitation}. They entail expanding the spectral sum \eqref{bigsumfield} according to
\begin{equation}\label{dressedexp}
\left\langle  \sigma( x,t) \sigma( 0,0)  \right\rangle={\cal D}^2+\sum_{n\geq 1}\mathfrak{S}^{\rm dr}_n\,,
\end{equation}
with
\begin{equation}\label{sdr}
\sum_{m=1}^n\mathfrak{S}^{\rm dr}_m=\sum _{ \substack{\pmb{\mu}\\ \exists \tau\text{ permutation of }\{1,...,N\},\\\,\# \{a\text{ s.t. }|I_a-J_{\tau(a)}|={\cal O}(L)\} \leq n}}\frac {\left| \left\langle \pmb{\mu} |\sigma( 0) |\pmb{\lambda}\right\rangle \right| ^{2}} {\left\langle \pmb{\lambda} \left| \pmb{\lambda} \right\rangle \left\langle \pmb{\mu}\right| \pmb{\mu}\right\rangle }e^{it\left( E\left( \pmb{\lambda}\right) -E\left( \pmb{\mu} \right) \right) +ix\left( P\left( \pmb{\mu}\right) -P\left( \pmb{\lambda}\right) \right) }\,.
\end{equation}

Loosely speaking, $\mathfrak{S}^{\rm dr}_n$ includes $n$ macroscopic particle-hole excitations and any number of microscopic particle-hole excitations. Since these configurations are not disjoint, one requires the expression \eqref{sdr} for a precise definition. In the low-density limit the full spectral sum truncates to $\mathfrak{S}^{\rm dr}_1$, and this expression is found to be finite and well-defined in the thermodynamic.

We note that these dressed particle-hole excitations are actually in principle what is computed in \cite{deNP15,deNP16,DNP18}. But therein the 'soft modes' contribute as a numerical factor that multiplies the form factor. Here, within the low density expansion, it is seen in \eqref{densityde} that these soft modes actually carry an $x$ and $t$ dependence as well. Namely we have a representation for $\mathfrak{S}_n^{\rm dr}$
\begin{equation}
\begin{aligned}
\underbrace{\int_{-\infty}^\infty...\int_{-\infty}^\infty}_{2n}&F_n^{x,t}(\lambda_1,\mu_1,...,\lambda_n,\mu_n)e^{it(E(\pmb{\lambda})-E(\pmb{\mu}))+ix(P(\pmb{\mu})-P(\pmb{\lambda}))}\\
&\qquad\qquad\qquad\times\rho(\lambda_1)\rho_h(\mu_1)...\rho(\lambda_n)\rho_h(\mu_n)\D{\lambda_1}\D{\mu_1}...\D{\lambda_n}\D{\mu_n}\,,
\end{aligned}
\end{equation}
where the 'dressed thermodynamic form factor' $F_n^{x,t}(\lambda_1,\mu_1,...,\lambda_n,\mu_n)$ carries a $x,t$ dependence coming from the soft modes summation, although the soft modes do not modify the energy and momentum of the state in the thermodynamic limit. This dependence emerges from the double poles that  lift to order $L^0$ some finite-size corrections. The function $F_1^{x,t}(\lambda,\mu)$ can be directly deduced from \eqref{densityde}.

From Sections \ref{which} and \ref{which2}, we conclude that the right expansion scheme of the spectral sum is an expansion in terms of dressed particle-hole excitations \eqref{dressedexp}, in the sense that the truncated spectral sums $\mathfrak{S}_n^{\rm dr}$ are finite and well-defined in the thermodynamic limit, and are smooth functions of the macroscopic excited rapidities. Moreover, as shown in \cite{granetessler20}, they admit a well-defined and uniform $1/c$ expansion without any spurious divergences in $L$ coming from $c$-dependent finite-size effects. These two properties are not satisfied by the 'bare' particle-hole excitations expansion.

\section{Summary and conclusion}
We computed the dynamical two-point function of the field and density operator averaged within a state with a small particle density ${\cal D}$, given by Equations \eqref{field} and \eqref{densityde}. They are valid for an arbitrary interaction strength $c>0$ and for all space and time -- hence one can deduce the spectral function and dynamical structure factor in the full momentum-frequency plane, in the same regime of small ${\cal D}$. This low-density limit is defined as the leading term in the correlation functions obtained by decomposing the form factors into partial fractions. 

Besides the explicit expressions obtained in the low-density regime, this work also provides interesting insights on the nature of states that contribute to the spectral sum, and on its possible expansions as detailed in Sections \ref{which} and \ref{which2}. The low-density regime is indeed naturally interpreted as a single dressed particle-hole excitation, i.e. a macroscopic particle-hole excitation with an arbitrary number of microscopic particle-hole excitations. In contrast, an integral representation in terms of 'bare' particle-hole excitations in the thermodynamic limit is found to be singular, in the sense that the 'thermodynamic form factor' for one-particle-hole excitations is non-zero only in the limit where the amplitude of the macroscopic excitation vanishes. Hence this work indicates that the right expansion of the spectral sum is in terms of dressed particle-hole excitations, as already suggested by the $1/c$ expansion developed in \cite{granetessler20}. Importantly, this dressing by microscopic particle-hole excitations also comes with an $x$ and $t$ dependence.

This PFD framework allows in principle for a computation of the next orders, which constitutes the most natural direction of improvement of this work. The fact that the next orders can indeed be computed with the PFD has been shown in \cite{GFE20} for the TFIM, that can be reformulated in terms of free fermions. In the Lieb-Liniger case, the interaction introduces technical but not fundamental difficulties, and we hope to be able to pursue this program in following works. Computing the next orders will also permit to assess the range of validity of the low-density approximation, which cannot be determined solely from this leading term. The next leading order will be particularly interesting, since it should contain certain terms of order $c^{-1}$ and $c^{-2}$ appearing in the $1/c$ expansion of these dynamical correlations at fixed root density \cite{granetessler20}. The comparison of this expansion in $\mathcal{D}$ at fixed $c$ with the expansion in $1/c$ at fixed $\mathcal{D}$ will permit us to assess their respective range of validity, in particular as a function of the parameter that is fixed.

Another interesting direction of development would be to apply the same framework to other integrable models such as the XXZ spin chain. However in this case there are also bound states ("strings") in the solutions of the Bethe equations, and one would have to study how to take them into account in the framework of the low-density expansion.

\paragraph{Acknowledgements}
We thank Fabian Essler and Jacopo De Nardis for helpful discussions. This work was
supported by the EPSRC under grant EP/S020527/1.

\end{document}